%%%%
%%%%    CHERN2.TEX                PLAIN TEX FILE 
%%%%
%%%%

\baselineskip 14pt plus 2pt

\font\llbf=cmbx10 scaled\magstep2
\font\lbf=cmbx10 scaled\magstep1

\def\ni{\noindent}

\def\ua{\underline a \,}      
\def\ub{\underline b \,}      
\def\uc{\underline c \,}      
\def\ud{\underline d \,}

      \def\bi{\bf i \,}
      \def\bj{\bf j \,}
\def\uk{\underline k \,}      \def\bk{\bf k \,}
      \def\bl{\bf l \,}
\def\um{\underline m \,}      \def\bm{\bf m \,}
\def\un{\underline n \,}      \def\bn{\bf n \,}

\def\uA{\underline A \,}      \def\ualpha{\underline\alpha \,}
\def\uB{\underline B \,}      \def\ubeta{\underline\beta \,}
\def\uC{\underline C \,}      \def\ugamma{\underline\gamma \,}
\def\uD{\underline D \,}      
\def\uE{\underline E \,}
\def\uF{\underline F \,}
\def\uX{\underline X \,}
\def\uY{\underline Y \,}

%%%%%%%%%%%%%%%%%%%%%%%%%%%%%%%%%%%%%%%%%%%%%%%%%%%%%%%%%%%%%%%%%%%%%%%%%%%
%%%%%%%%%%%%%%%%%%%%%%%%%%%%%%%%%%%%%%%%%%%%%%%%%%%%%%%%%%%%%%%%%%%%%%%%%%%

\ni
{\llbf On certain global conformal invariants and 3-surface twistors of 
initial data sets}
\par
\bigskip
\bigskip
\ni
{\bf L\'aszl\'o B. Szabados}\par
\bigskip
\ni
Research Institute for Particle and Nuclear Physics \par
\ni
H-1525 Budapest 114, P.O.Box 49, Hungary \par
\ni
E-mail: lbszab@rmki.kfki.hu \par
\bigskip
\bigskip

\ni
The Chern--Simons functionals built from various connections determined 
by the initial data $h_{\mu\nu}$, $\chi_{\mu\nu}$ on a 3-manifold $\Sigma$ 
are investigated. First it is shown that for asymptotically flat data sets 
the logarithmic fall off for $h_{\mu\nu}$ and $r\chi_{\mu\nu}$ is the 
necessary and sufficient condition of the existence of these functionals. 
The functional $Y_{(k,l)}$, built in the vector bundle corresponding to the 
irreducible representation of $SL(2,{\bf C})$ labelled by $(k,l)$, is shown 
to be determined by the Ashtekar--Chern--Simons functional and its complex 
conjugate. $Y_{(k,l)}$ is conformally invariant precisely in the $l=k$ 
(i.e. tensor) representations. An unexpected connection with twistor theory 
is found: $Y_{(k,k)}$ can be written as the Chern--Simons functional built 
from the 3-surface twistor connection, and the not identically vanishing 
spinor parts of the 3-surface twistor curvature are given by the variational 
derivatives of $Y_{(k,k)}$ with respect to $h_{\mu\nu}$ and $\chi_{\mu\nu}$. 
The time derivative $\dot Y_{(k,k)}$ of $Y_{(k,k)}$ is another global 
conformal invariant of the initial data set, and for vanishing $\dot Y
_{(k,k)}$, in particular for all Petrov III. and N spacetimes, the 
Chern--Simons functional is a conformal invariant of the whole spacetime. 
\bigskip
\bigskip

\ni
{\lbf 1. Introduction}\par
\bigskip
\ni
In a recent joint paper with Robert Beig the conformal invariant $Y[h_{\mu
\nu}]$ of Chern and Simons [1], defined for closed orientable Riemannian 
3-manifolds $(\Sigma,h_{\mu\nu})$, was generalized for triples $(\Sigma,
h_{\mu\nu},\chi_{\mu\nu})$, where $\chi_{\mu\nu}$ is a symmetric tensor 
field [2]. (For the sake of simplicity we call such a triple an initial 
data set and $\chi_{\mu\nu}$ the extrinsic curvature even if we don't 
use any field equation, not even any constraints for $h_{\mu\nu}$ and 
$\chi_{\mu\nu}$; and even if $(\Sigma,h_{\mu\nu},\chi_{\mu\nu})$ is not 
assumed to be imbedded in any spacetime.) Similarly to $Y[h_{\mu\nu}]$, 
the new $Y_0[h_{\mu\nu},\chi_{\mu\nu}]$ was defined as the integral of 
the Chern--Simons 3-form, built from the connection on an appropriate 
vector bundle over $\Sigma$, modulo $16\pi^2$. In the former case the 
connection was the Levi-Civita connection determined by $h_{\mu\nu}$ on 
the tangent bundle of $\Sigma$, whilst in the latter it was the real Sen 
connection determined by $h_{\mu\nu}$ and $\chi_{\mu\nu}$ on a 
trivializable Lorentzian vector bundle, i.e. the vector bundle constructed 
by the vector representation of $SL(2,{\bf C})$, over $\Sigma$. If 
$\Sigma$ is a spacelike hypersurface in a Lorentzian spacetime, then this 
vector bundle is just the spacetime tangent bundle pulled back to $\Sigma$. 
Thus, roughly speaking, we retain the four dimensional Lorentzian character 
of the geometry of the initial data set infinitesimally, i.e. at the level 
of tangent spaces, even in a 3+1 decomposition of spacetime. $Y[h_{\mu\nu}]$ 
is known to be invariant with respect to conformal rescalings of the 
3-metric $h_{\mu\nu}$, and $Y_0[h_{\mu\nu},\chi_{\mu\nu}]$ turned out to be 
invariant with respect to changes of $h_{\mu\nu}$ and $\chi_{\mu\nu}$ 
corresponding to {\it spacetime} conformal rescalings. The functional 
derivative of $Y[h_{\mu\nu}]$ with respect to $h_{\mu\nu}$ is known to be 
the Cotton--York tensor, and hence the stationary points of $Y[h_{\mu\nu}]$ 
are the locally conformally flat Riemannian 3-manifolds. The variational 
derivatives of $Y_0[h_{\mu\nu},\chi_{\mu\nu}]$ yield two symmetric 
trace-free tensor fields, $B_{\mu\nu}$ and $H_{\mu\nu}$, whose vanishing 
characterizes the local isometric imbeddability of $(\Sigma,h_{\mu\nu},\chi
_{\mu\nu})$ into some conformal Minkowski spacetime. $H_{\mu\nu}$ is the 
conformal magnetic curvature, while $B_{\mu\nu}$ is the natural 
generalization of the Cotton--York tensor for non-vanishing $\chi_{\mu
\nu}$. The nontriviality of these invariants is shown by a result of 
Meyerhoff [3], namely that in the Riemannian case for certain hyperbolic 
manifolds $Y[h_{\mu\nu}]$ takes values which are dense on the circle $S^1=
{\bf R}$ modulo $16\pi^2$. In [2] the analogous functional $Y_\pm[h_{\mu
\nu},\chi_{\mu\nu}]$ based on the {\it complex} self-dual/anti-self-dual 
Ashtekar connection, i.e. based on the bundle constructed by the 
self-dual/anti-self-dual representation of $SL(2,{\bf C})$, was also 
considered and was shown {\it not} to be conformally invariant. In fact, 
the stationary points of $Y_\pm[h_{\mu\nu},\chi_{\mu\nu}]$ are precisely 
those data sets that can be locally isometrically imbedded into the 
Minkowski spacetime. \par
       From physical points of view it would be desirable to be able to 
define the conformal invariant $Y_0:=Y_0[h_{\mu\nu},\chi_{\mu\nu}]$ for 
asymptotically flat initial data sets too, because these data sets are 
thought to represent the gravitational field of localized objects. In 
particular, this may provide a useful tool in studying the structure of 
spacelike and null infinity (see e.g. [4,5]), since it serves as a natural 
foliation of the space of initial data for the conformally equivalent 
spacetimes. As we mentioned above, the critical points of $Y_0$ are 
precisely those data sets that can be imbedded into some conformally flat 
spacetime. But, as Tod proved [6], this imbeddability is equivalent to the 
complete integrability of the 3-surface twistor equation. Taking into 
account the conformal invariance of $Y_0$, one might conjecture that there 
is a hidden connection between our previous construction and various 
3-surface twistor concepts. In particular $Y_0$ might be a functional of 
the 3-surface twistor connection. Furthermore, it might be interesting even 
from twistor theoretical points of view to clarify the properties of the 
Chern--Simons functional built from the 3-surface twistor connection. Since 
the functionals $Y_0$, $Y_\pm:=Y_\pm[h_{\mu\nu},\chi_{\mu\nu}]$ have different 
conformal properties, the question arises whether new nontrivial invariants 
can be obtained by considering other vector bundles, i.e. representations 
of $SL(2,{\bf C})$, or not. (Recently another interesting connection was 
introduced in the canonical description of general relativity, the 
so-called Barbero connection [7,8]. Although the Barbero--Chern--Simons 
functional has several interesting properties, e.g. it is conformally 
invariant precisely in the tensor representations for any real value of the 
Barbero--Immirzi parameter, we will not consider that in the present paper.) 
Further interesting issue is the question of the time evolution of these 
functionals, i.e. how they change as the function of time if the data is 
evolved in time (e.g. by Einstein's field equations). \par
        In the present paper we investigate (further) the properties of 
$Y_0$, $Y_\pm$, and the Chern--Simons functional built from the 3-surface
twistor connection. In the first two subsections we review the main points 
of the construction of $Y_0$ and discuss how $h_{\mu\nu}$ and $\chi_{\mu
\nu}$ determine it uniquely. (This issue was not exhaustively discussed in 
[2]. Moreover, we improve several points of the presentation and correct 
some minor numerical errors.) We define $Y_0$ for asymptotically flat 
initial data sets by determining the weakest possible fall-off conditions 
for $h_{\mu\nu}$ and $\chi_{\mu\nu}$. Since however the Chern--Simons 
functional is defined in the tetrad rather than the metric theory, a new 
technique was needed to determine the asymptotic behaviour of the initial 
data. This technique can also be used to derive the fall-off and asymptotic 
gauge conditions in the canonical analysis of the (tetrad or triad) general 
relativity. We will see that the weakest possible fall-off conditions are 
much weaker than those coming from general relativity, namely logarithmic 
fall-off for $h_{\mu\nu}$ and $r\chi_{\mu\nu}$, and hence the Chern--Simons 
conformal invariant is well defined for asymptotically flat initial data 
sets for Einstein's theory. These fall-off conditions ensure the existence 
of the Ashtekar--Chern--Simons functional as well. For later use (especially 
in the twistorial approach) the construction is rewritten in the spinor 
representation in subsection 2.3. In this representation the real Sen-- and 
the complex Ashtekar--Chern--Simons functionals can be treated 
simultaneously. Finally, we clarify how the Chern--Simons functional 
depends on the representation of the structure group by showing that 
the construction, based on a general finite dimensional representation, 
doesn't give anything new, that is simply a combination of the 
Ashtekar--Chern--Simons functional and its complex conjugate, or, 
equivalently, the Sen--Chern--Simons and the Ashtekar--Chern--Simons 
functionals. The conformally invariant functionals correspond precisely to 
the tensor representations. \par
      In section three the potential relation to twistor theory will be 
clarified. In subsection 3.1 the unitary spinor forms of the Sen operator 
and the concept of 3-surface twistors will be reviewed. Although most of 
that subsection is essentially a review (mainly to fix the notations, to
present the tools for the next subsection and to retain the coherence and 
readability of the paper), it contains several new elements, e.g. the 
unitary spinor form of the full Ricci and Bianchi identities and the tensors 
$H_{\mu\nu}$, $B_{\mu\nu}$, too. Then, in subsection 3.2, we calculate the 
3-surface twistor connection and curvature explicitly, and show that the 
tensors $H_{\mu\nu}$, $B_{\mu\nu}$ above represent the non-vanishing 
components of the 3-surface twistor curvature. Thus $H_{\mu\nu}$ and $B_{\mu
\nu}$ have natural twistorial interpretation. Then the Sen--Chern--Simons 
functional will be shown to be just twice the Chern--Simons functional built 
from the 3-surface twistor connection. \par
       Section four is devoted to the problem of time evolution of the 
Chern--Simons functionals. We derive a formula by means of which we can 
compare these functionals on arbitrary two spacelike hypersurfaces and 
determine the conditions of their hypersurface--independence. These 
conditions are satisfied for a large class of algebraically general and 
special spacetimes, including all the Petrov III. and N. type metrics, 
yielding two new global invariants for these spacetimes. One of them is a 
global {\it conformal invariant}. Finally, we discuss the properties of the 
imaginary part of the Ashtekar--Chern--Simons functional, a proposal for 
the natural time variable in cosmological spacetimes, and we will see that 
${\rm Im}Y_\pm$ is monotonic only for a very limited class of spacetimes. 
Finally we calculate the Chern--Simons functional for the general closed 
homogeneous Bianchi cosmologies with simply-transitive group actions, and, 
in particular, for the vacuum Kasner, the general Robertson--Walker and 
the special anisotropic Barrow solutions. This is the only point where 
Einstein's equations are used in the present paper. These examples show the 
usefulness and the nontriviality of the generalizations $Y_0$, $Y_\pm$ of 
the (Riemannian) conformal invariant of Chern and Simons for initial data 
sets. \par
      Our general spinor--twistor reference is [9], and that of differential
geometry is [10]. In particular, the wedge product of forms is defined to be 
the anti-symmetric part of the tensor product, the signature of the spacetime 
and spatial metrics is (+ -- -- --) and (-- -- --), respectively. The 
curvature $F^a{}_{b\mu\nu}$ of a covariant derivative ${\cal D}_\mu$ is 
defined by $-F^a{}_{b\mu\nu}X^bv^\mu w^\nu:=v^\mu{\cal D}_\mu(w^\nu{\cal D}
_\nu X^a)-w^\mu{\cal D}_\mu(v^\nu{\cal D}_\nu X^a)-[v,w]^\mu{\cal D}_\mu 
X^a$. The Ricci tensor is $R_{\mu\nu}:=R^\rho{}_{\mu\rho\nu}$, and the 
curvature scalar is the contraction of $R_{\mu\nu}$ and the metric. We 
changed our previous notations slightly. We use several types of indices, 
both abstract and concrete (name) indices, whose range will be explained 
when they appear first. \par
\bigskip

\ni
{\lbf 2. The Chern--Simons functional of asymptotically flat initial 
data sets}\par
\bigskip
\ni
{\bf 2.1 The general Chern--Simons functional}\par
\medskip
\ni
Let $\Sigma$ be a connected orientable 3-manifold, which is asymptotically 
Euclidean in the sense that for some compact set $K\subset\Sigma$ the 
complement $\Sigma-K$ is diffeomorphic to ${\bf R}^3-B$, where $B$ is a 
closed ball in ${\bf R}^3$. This complement represents the `asymptotic end' 
of $\Sigma$.\footnote{*}{For the sake of simplicity we assume that $\Sigma$ 
has one asymptotic end. It is obvious how to generalize $\Sigma$ to have 
more than one such ends.} Greek indices from the second half of the Greek 
alphabet, e.g. $\mu,\nu,...$, will be abstract tensor indices referring to 
$\Sigma$ in general, but in the present subsection they denote concrete 
coordinate indices, too. Let $G$ be a Lie group, ${\cal G}$ its Lie algebra, 
and $\pi:P\rightarrow\Sigma$ a trivializable principal bundle over $\Sigma$ 
with structure group $G$. Let ${\bf E}^a$ a $k$ dimensional vector space 
over {\bf K}={\bf R} or {\bf C}, $\rho:G\rightarrow GL({\bf E}^a)$ a linear 
representation of $G$ on ${\bf E}^a$, $\rho_\ast:{\cal G}\rightarrow gl({\bf 
E}^a)$ the corresponding representation of the Lie algebra, and $\pi:E^a(
\Sigma)\rightarrow\Sigma$ the associated (trivializable) vector bundle. 
Because of the trivializability $E^a(\Sigma)$ admits $k$ global cross 
sections, $E^a_{\ua}$, ${\ua}=1,...,k$, such that at each point $p\in
\Sigma$ $\{E^a_{\ua}\vert_p\}$ spans the fibre $\pi^{-1}(p)\subset 
E^a(\Sigma)$. Such a collection of cross sections of $E^a(\Sigma)$ will be 
called a global frame field. Thus small Roman indices are abstract `internal 
bundle' indices, while the underlined small Roman indices are name indices. 
The global cross section $\sigma:\Sigma
\rightarrow P$ of the principal bundle defines a transformation $\rho
\circ\sigma$ of the global frame fields (`globally defined local gauge 
transformations'); i.e. it is a $k\times k$ matrix valued function 
$\Lambda^{\ua}{}_{\ub}$ on $\Sigma$ acting on a global frame field as 
$E^a_{\ua}\mapsto E^a_{\ua}\Lambda^{\ua}{}_{\ub}$. \par
       Any connection on $P$ defines a connection on $E^a(\Sigma)$, whose 
connection coefficients with respect to a global frame field form a $\rho
_\ast({\cal G})\subset gl(k,{\bf K})$-valued 1-form $A^{\ua}_{\mu{\ub}}$ on 
$\Sigma$, and the curvature of this connection is the $\rho_\ast({\cal 
G})$-valued 2-form $-F^{\ua}{}_{{\ub}\mu\nu}:=\partial_\mu A^{\ua}
_{\nu\ub}-\partial_\nu A^{\ua}_{\mu\ua}+A^{\ua}_{\mu\uc}A^{\uc}_{\nu\ub}-
A^{\ua}_{\nu\uc}A^{\uc}_{\mu\ub}$. Using the matrix notation for the 
underlined (`internal name') indices, the Chern--Simons functional is well 
known to be defined by 

$$
Y[A]:=\int_\Sigma {\rm Tr}\Bigl(F_{[\mu\nu}A_{\rho]}+{2\over3}A_{[\mu}A_\nu 
A_{\rho]}\Bigr). \eqno(2.1.1)
$$
\ni
To ensure the existence of this integral we must impose certain fall-off 
conditions on the connection coefficients. \par
      Let $(r,\theta,\phi)$ be the standard polar coordinates on $\Sigma-K
\approx{\bf R}^3-B$, let $\{E^a_{\ua}\}$ be a fixed global frame field 
and determine the fall-off condition for $A^{\ua}_{\mu{\ub}}$ with
respect to these coordinates and global frame field, implied by the 
existence of $Y[A]$. Since on the asymptotic end the integrand of 
(2.1.1) takes the form $-2{\rm Tr}(A_\mu\partial_\nu A_\rho+{1\over3}A_\mu 
A_\nu A_\rho)\epsilon^{\mu\nu\rho}{\rm d}r{\rm d}\theta{\rm d}\phi$, where 
$\mu,\nu,...=r,\theta,\phi$ and $\epsilon^{\mu\nu\rho}=\epsilon_{\mu\nu
\rho}$ is the alternating Levi-Civita symbol, it seems natural to impose 
the following fall-off conditions 

$$\eqalign{
A^{\ua}_{r{\ub}}(r,\theta,\phi)=&{A^{\ua}{}_{\ub}(\theta,\phi)\over r^{\rm 
a}}+o\bigl({1\over r^{\rm a}}\bigr), \cr
A^{\ua}_{\tau{\ub}}(r,\theta,\phi)=&{A^{\ua}_{\tau\ub}(\theta,\phi)\over 
r^{\rm b}}+o\bigl({1\over r^{\rm b}}\bigr), \hskip 20pt 
  \tau=\theta,\phi,\cr
{\rm for\,\,some}\hskip 20pt&{\rm a}+{\rm b}>1,\hskip 10pt {\rm b}>0,\cr} 
\eqno(2.1.2)
$$
\ni
where a function $f(r)$ is said to behave at infinity like $o(r^{-{\rm a}})$ 
if $\lim_{r\rightarrow\infty}(r^{\rm a}f(r))=0$. If the $A^{\ua}_{r{\ub}}$ 
component has $1/r$ fall-off, then the logarithmic fall-off for the 
tangential components $A^{\ua}_{\tau{\ub}}$ is the necessary and sufficient 
condition of the existence of (2.1.1). If therefore ${\cal A}$ denotes the 
set of all the connection 1-forms on $E^a(\Sigma)$ satisfying the fall-off 
condition (2.1.2) then $Y:{\cal A}\rightarrow{\bf K}$ becomes well defined. 
If $A^{\ua}_{\mu\ub}(u)$ is a smooth 1 parameter family of 
connections in ${\cal A}$ then the derivative of the Chern--Simons 
functional $Y[A(u)]$ with respect to $u$, i.e. the `variation' of $Y[A]$, 
is $\delta Y[A]:=({{\rm d}\over{\rm d}u}Y[A(u)])\vert_{u=0}=2\int_\Sigma(
{\rm Tr}F_{[\mu\nu}\delta A_{\rho]}+\partial_{[\mu}({\rm Tr}A_\nu\delta 
A_{\rho]}))$, where $\delta A^{\ua}_{\mu\ub}:=({{\rm d}\over{\rm d}u}
A^{\ua}_{\mu\ub}(u))\vert_{u=0}$, the `variation' of the connection 1-form. 
Thus the fall-off condition (2.1.2) ensure the functional differentiability 
of $Y[A]$ with respect to the connection 1-form, and the functional 
derivative is essentially the curvature. \par
       Under a gauge transformation $\Lambda:\Sigma\rightarrow\rho(G)$ the 
connection 1-form transforms as $A^{\ua}_{\mu\ub}\mapsto A^{\prime\ua}
_{\mu\ub}:=\Lambda_{\ud}{}^{\ua}(A^{\ud}_{\mu\uc}\Lambda^{\uc}{}_{\ub}
+\partial_\mu\Lambda^{\ud}{}_{\ub})$, where $\Lambda_{\ua}{}^{\ub}$ is 
defined by $\Lambda^{\ua}{}_{\uc}\Lambda_{\ub}{}^{\uc}=\delta^{\ua}_{\ub}$. 
Thus the gauge transformations preserve the fall-off properties of the 
connection 1-forms, i.e. they don't take a connection 1-form $A^{\ua}
_{\mu\ub}$ out of ${\cal A}$, if 

$$
\Lambda^{\ua}{}_{\ub}(r,\theta,\phi)={}_0\Lambda^{\ua}{}_{\ub}+{\Lambda
^{\ua}{}_{\ub}(\theta,\phi)\over r^{\rm c}}+o({1\over r^{\rm c}}), \hskip 
20pt  {\rm c}\geq{\rm max}\{{\rm a}-1,{\rm b}\}, \eqno(2.1.3)
$$
\ni
where ${}_0\Lambda^{\ua}{}_{\ub}$ is a constant $\rho(G)$-matrix. 
Under these gauge transformations $Y[A]$ transforms as 

$$\eqalign{
Y[A]-Y[A^\prime]&={2\over3}\int_\Sigma\Lambda_{\uk}{}^{\ua}\bigl(\partial
 _\mu\Lambda^{\um}{}_{\ua}\bigr)\Lambda_{\um}{}^{\ub}\bigl(\partial_\nu
 \Lambda^{\un}{}_{\ub}\bigr)\Lambda_{\un}{}^{\uc}\bigl(\partial_\rho\Lambda
 ^{\uk}{}_{\uc}\bigr){1\over3!}\delta^{\mu\nu\rho}_{\sigma\tau\omega}+\cr
&+2\int_\Sigma\partial_\mu\Bigl(A^{\ua}_{\nu\ub}\Lambda_{\ua}{}^{\uc}\bigl(
 \partial_\rho\Lambda^{\ub}{}_{\uc}\bigr)\Bigr){1\over3!}\delta^{\mu\nu\rho}
 _{\sigma\tau\omega},\cr}\eqno(2.1.4)
$$
\ni
where the second term, the integral of an exact 3-form, vanishes as a 
consequence of the fall-off conditions. For small gauge transformations 
(i.e. for gauge transformations $\Lambda:\Sigma\rightarrow\rho(G)$ 
homotopic to the identity transformation) the integrand of the first term 
in (2.1.4) is also exact, and hence the right hand side of (2.1.4) is zero, 
but for large gauge transformations (i.e. which are not small) the right 
hand side is $16\pi^2N$ for some {\it integer} $N$ depending on the 
homotopy class of $\Lambda:\Sigma\rightarrow\rho(G)$. This implies that 
$Y[A]$ modulo $16\pi^2$ is gauge invariant, and if $Y[A]$ is complex valued 
then its imaginary part ${\rm Im}\,Y[A]$ in itself is gauge invariant. If 
$\psi:\Sigma\rightarrow\Sigma$ is any smooth proper map then $Y[\psi^*A]=
{\rm deg}(\psi)Y[A]$, where ${\rm deg}(\psi)$ is the degree of $\psi$ [11]. 
Since however ${\rm deg}(\psi)$ is one for orientation preserving 
diffeomorphisms, $Y[A]$ is invariant with respect to them. \par
\bigskip

\ni
{\bf 2.2 The Sen--Chern--Simons functional of initial data sets}\par
\medskip
\ni
Let $\pi:L\rightarrow\Sigma$ be a trivializable principal fiber bundle over 
$\Sigma$ with structure group $SO_0(1,3)$, the connected component of the 
Lorentz group $O(1,3)$, $\rho_0$ its defining representation on the four 
dimensional real vector space ${\bf V}^a$, and $V^a(\Sigma)$ the associated 
vector bundle and $V_a(\Sigma)$ its dual vector bundle. Let $E^a_{\ua}$, 
${\ua}=0,...,3$, be a global frame field in $V^a(\Sigma)$ with given `space' 
and `time' orientation, and let $\vartheta^{\ua}_a$ be the dual global 
frame field in the dual bundle. Thus small Roman indices are abstract 
`internal' Lorentzian indices, i.e. they refer to the Lorentzian vector 
bundle, while underlined small Roman indices are Lorentzian name indices. 
If $\eta_{\ua\ub}:={\rm diag}(1,-1,-1,-1)$, then $g_{ab}:=\eta_{\ua\ub}
\vartheta^{\ua}_a\vartheta^{\ub}_b$ is a Lorentzian fibre metric on $V^a
(\Sigma)$, and $E^a_{\ua}$ becomes a $g_{ab}$-orthonormal global frame 
field. The global cross sections of $L$ define global gauge transformations 
taking $g_{ab}$-orthonormal global frame fields into $g_{ab}$-orthonormal 
global frame fields. $g_{ab}$ identifies $V^a(\Sigma)$ with its dual $V_a
(\Sigma)$. Let $\Theta:T\Sigma\rightarrow V^a(\Sigma):(p,v^\mu)\mapsto(p,
v^\mu\Theta^a_\mu)$ be an imbedding of $T\Sigma$ into $V^a(\Sigma)$ such 
that $h_{\mu\nu}:=\Theta^a_\mu\Theta^b_\nu g_{ab}$ is a negative definite 
metric on $T\Sigma$. One can raise and lower the indices of $\Theta^a_\mu$ 
by $h^{\mu\nu}$ and $g_{ab}$, respectively, and e.g. $\Theta^\mu_a$ defines 
an imbedding of $T^\ast\Sigma$ into $V_a(\Sigma)$. Let $t^a$ be the section 
of $V^a(\Sigma)$ which is orthogonal to $\Theta(T\Sigma)$, i.e. $v^\mu\Theta
^a_\mu t_a=0$ for any section $v^\mu$ of $T\Sigma$, and has unit norm with 
respect to $g_{ab}$. The orientation of $t^a$ is chosen to be compatible 
with the `time' orientation of the global frame fields, e.g. to be `future' 
directed. Then $P^a_b:=\delta^a_b-t^at_b$ is the projection of the fibre 
$V^a_p$ onto $\Theta(T_p\Sigma)$ for any $p\in\Sigma$, and hence any 
section $X^a$ of $V^a(\Sigma)$ can be decomposed in a unique way as $X^a=
Nt^a+N^a$, where $N^a=P^a_bN^b$ is called the shift and $N$ is the lapse 
part of $X^a$. This decomposition defines a vector bundle 
isomorphism between the Whitney sum of the trivial line bundle over 
$\Sigma$ and $T\Sigma$, and the Lorentzian vector bundle: $i:(\Sigma
\times{\bf R})\oplus T\Sigma\rightarrow V^a(\Sigma): (p,(N,N^\mu))\mapsto(p,
Nt^a+N^\mu\Theta^a_\mu)$. Any $h_{\mu\nu}$-orthonormal frame field $e^\mu
_{\bi}$, ${\bi}=1,2,3$, in $T\Sigma$ defines a $g_{ab}$-orthonormal global 
frame field $\{t^a,e^a_{\bi}\}$ in $V^a(\Sigma)$ by $e^a_{\bi}:=e^\mu_{\bi}
\Theta^a_\mu$. Such a frame field in $V^a(\Sigma)$ will be said to be 
compatible with the imbedding $\Theta$, and the set of all such 
$\Theta$-compatible frame fields defines a reduction $SO_0(1,3)\rightarrow 
SO(3)$ of the gauge group (`time gauge'). Since the quotient $SO_0(1,3)/
SO(3)$ is homeomorphic to ${\bf R}^3$, there always exist {\it small} gauge 
transformations taking a global frame field into a $\Theta$-compatible 
frame field [2]. \par
    Any connection on $\pi:L\rightarrow\Sigma$ defines a $g_{ab}$-compatible 
covariant derivative ${\cal D}_\mu$ on $V^a(\Sigma)$, which can be 
characterized completely by its action on pointwise independent sections 
of $V^a(\Sigma)$, e.g. by $\chi_{\mu a}:={\cal D}_\mu t_a$ and ${\cal D}
_\mu(e^\nu_{\bi}\Theta^b_\nu)$. We call ${\cal D}_\mu$ the real Sen 
connection on $V^a(\Sigma)$ if the next three conditions are satisfied:
\item{i.}   ${\cal D}_\mu g_{ab}=0$, 
\item{ii.}  $\chi_{\mu\nu}:=\bigl({\cal D}_\mu t_a\bigr)\Theta^a_\nu
            =\chi_{(\mu\nu)}$, 
\item{iii.} ${\cal D}_\mu\bigl(e^\nu_{\bi}\Theta^b_\nu\bigr)P^a_b=\bigl(
            D_\mu e^\nu_{\bi}\bigr)\Theta^a_\nu$, where $D_\mu$ is the 
	     Levi-Civita covariant derivative on $T\Sigma$ determined by 
	     the metric $h_{\mu\nu}$.
\par
\ni
For fixed bundle isomorphism $i$ and tensor fields $h_{\mu\nu}$ and 
$\chi_{\mu\nu}$ these conditions uniquely determine the derivative ${\cal 
D}_\mu$. The ${\cal D}_\mu$--derivative of the section $X^a=Nt^a+N^a$ is 
${\cal D}_\mu X^a=(D_\mu N)t^a+({\cal D}_\mu N^b)P^a_b+(\chi_\mu{}^at_b-
t^a\chi_{\mu b})X^b$. Thus it seems useful to define the action of the 
Levi-Civita derivative $D_\mu$ on sections $v^a$ of $V^a(\Sigma)$ satisfying 
$v^a=v^bP^a_b$ by $D_\mu v^a:=D_\mu(v^b\Theta^\nu_b)\Theta^a_\nu$ (see 
requirement iii. above), and then to extend its action to any section of 
$V^a(\Sigma)$ by demanding $D_\mu t_a=0$, since then both ${\cal D}_\mu$ 
and $D_\mu$ would be defined on the same vector bundle and one could compare 
them. Their difference is $({\cal D}_\mu -D_\mu)X^a=(\chi_\mu{}^at_b-t^a\chi
_{\mu b})X^b$. For the sake of later convenience let us introduce $V_{\mu
\nu\rho\omega}:=\chi_{\mu\rho}\chi_{\nu\omega}-\chi_{\mu\omega}\chi_{\nu
\rho}$ and its traces $V_{\mu\nu}:=V^\rho{}_{\mu\rho\nu}$ and $V=V^\rho{}
_\rho$. They have all the algebraic symmetries of the Riemann and Ricci 
tensors in three dimensions. The curvature of ${\cal D}_\mu$ has the form 
$F^{ab}{}_{\mu\nu}=\Theta^a_\rho\Theta^b_\omega(R^{\rho\omega}{}_{\mu\nu}+
V^{\rho\omega}{}_{\mu\nu})+(t^a\Theta^b_\omega-t^b\Theta^a_\omega)
(D_\mu\chi_\nu{}^\omega-D_\nu\chi_\mu{}^\omega)$. Here $R^\rho{}_{\omega\mu
\nu}$ is the curvature tensor of $(\Sigma,h_{\mu\nu})$. The `Ricci part' of 
the curvature is $F^{ab}{}_{\mu\nu}\Theta^\nu_b=\Theta^a_\rho(R^\rho{}_\mu+
V^\rho{}_\mu)-t^a(D_\nu\chi^\nu{}_\mu-D_\mu\chi)$, which, contrast to 
Riemannian 3-manifolds, doesn't determine the full curvature of ${\cal D}
_\mu$. What remains undetermined is the term that can be represented by 
$H_{\mu\nu}:=-\varepsilon_{\rho\omega(\mu}D^\rho\chi^\omega{}_{\nu)}$. 
If $(M,g_{ab})$ is a Lorentzian spacetime and $\theta:\Sigma\rightarrow M$ 
is an imbedding such that $\theta(\Sigma)$ is spacelike, then $V^a(\Sigma)$ 
can be identified with the pull back to $\Sigma$ of the spacetime tangent 
bundle $TM$ along $\theta$ and $\Theta^a_\mu$ is the differential of 
$\theta$. The Sen connection ${\cal D}_\mu$ introduced here is $\Theta^a
_\mu{\cal D}_a$, the pull back to $\Sigma$ of the derivative ${\cal D}_a:=
P^b_a\nabla_b$ of Sen [12], and its curvature is just the pull back to 
$\Sigma$ of the spacetime curvature tensor: $F^a{}_{b\mu\nu}={}^{(4)}R^a{}
_{bcd}\Theta^c_\mu\Theta^d_\nu$. Then the tensor $H_{\mu\nu}$ becomes the 
pull back to $\Sigma$ of the magnetic part $H_{ab}:={1\over2}\varepsilon
_{ac}{}^{ef}C_{efbd}t^ct^d$ of the spacetime Weyl tensor. (Note that we use 
the convention in which the relation between the three and four dimensional 
volume forms is $\varepsilon_{abc}=\varepsilon_{abcd}t^d$.) On the other 
hand, in general the electric part of the spacetime Weyl tensor, $E_{ab}:=
C_{acbd}t^ct^d$, cannot be expressed by the geometric data on $\Sigma$. It 
contains the spatial-spatial part of the spacetime Einstein tensor and the 
spacetime curvature scalar too: $E_{ab}=-(R_{ab}+V_{ab}-{1\over2}{}^{(4)}
G_{cd}P^c_aP^d_b)+{1\over4}h_{ab}(R+V+{2\over3}{}^{(4)}R)$. The `constraint 
parts' of the spacetime Einstein tensor are ${}^{(4)}G_{ab}t^at^b=-{1\over2}
(R+V)$ and ${}^{(4)}G_{ab}t^aP^b_c=-D_a(\chi^a{}_c-\chi\delta^a_c)$. \par
         The connection coefficients of ${\cal D}_\mu$ with respect to 
any pair of dual global frame fields are $\Gamma^{\ua}_{\mu{\ub}}:=\vartheta
^{\ua}_a{\cal D}_\mu E^a_{\ub}$, and, following the general prescription 
of the previous subsection, we can form the Chern--Simons functional built 
from the real Sen connection. This $Y[\Gamma^{\ua}{}_{\ub}]$ will be called 
the Sen--Chern--Simons functional. Since the difference of the Chern--Simons 
3-form in one gauge and in another gauge obtained by a small gauge 
transformation is an exact 3-form, the Sen--Chern--Simons functional can 
always be calculated in the time gauge. In the frame field compatible 
with the imbedding $\Theta$ the connection coefficients are $\Gamma^0_{\mu
{\bj}}=-\chi_{\mu\nu}e^\nu_{\bj}$ and $\Gamma^{\bi}_{\mu{\bj}}=\zeta^{\bi}
_\nu D_\mu e^\nu_{\bj}$, the Ricci rotation coefficients of $D_\mu$. Here 
$\{\zeta^{\bi}_\nu\}$ is the global frame field in $T^*\Sigma$ dual to $\{
e^\nu_{\bi}\}$. The curvature 2-form, also in the time gauge, is given by 
$F^{\bi}{}_{{\bj}\mu\nu}=\zeta^{\bi}_\rho e^\omega_{\bj}(R^\rho{}_{\omega
\mu\nu}+V^\rho{}_{\omega\mu\nu})$ and $F^0{}_{{\bj}\mu\nu}=e^\omega_{\bj}
(D_\mu\chi_{\nu\omega}-D_\nu\chi_{\mu\omega})$. Since however these 
expressions depend only on the triad field $\{e^\mu_{\bi}\}$ and the tensor 
field $\chi_{\mu\nu}$ and independent of the vector bundle isomorphism $i$ 
(or even the imbedding $\Theta$), the Sen--Chern--Simons functional will 
be completely determined by $\{e^\mu_{\bi}\}$ and $\chi_{\mu\nu}$. Finally, 
since $Y[e^\mu_{\bi},\chi_{\mu\nu}]$ modulo $16\pi^2$ is gauge invariant, 
it is a functional only of $h_{\mu\nu}$ and $\chi_{\mu\nu}$ and will be 
denoted by $Y_0[h_{\mu\nu},\chi_{\mu\nu}]$.\par
      Next determine the fall-off properties of $h_{\mu\nu}$ and $\chi_{\mu
\nu}$ implied by the general fall-off conditions (2.1.2). Let ${}_0h_{\mu
\nu}$ be a fixed negative definite metric on $\Sigma$ such that the 
asymptotic end $\Sigma-K$, together with the restriction of ${}_0h_{\mu\nu}$ 
to $\Sigma-K$, is isometric to the standard flat geometry on ${\bf R}^3-B$. 
Let $\{{}_0\zeta^{\bi}_\mu,{}_0e^\mu_{\bi}\}$ be a ${}_0h_{\mu
\nu}$-orthonormal dual frame field which is {\it constant} on $\Sigma-K$, 
i.e. ${}_0D_\mu{}_0\zeta^{\bi}_\nu=0$, and let the orientation of these 
frame fields be chosen to be that of the $h_{\mu\nu}$-orthonormal `physical' 
frames $\{\zeta^{\bi}_\mu,e^\mu_{\bi}\}$. Since both $\{{}_0\zeta^{\bi}
_\mu\}$ and $\{\zeta^{\bi}_\mu\}$ are bases with the same orientation, they 
can be combined from each other, i.e. for some globally defined $GL(3,{\bf 
R})$-valued function $\Phi_{\bj}{}^{\bi}$ on $\Sigma$ with positive 
determinant we have $\zeta^{\bi}_\mu={}_0\zeta^{\bj}_\mu\Phi_{\bj}{}^{\bi}$. 
If $\Phi^{\bi}{}_{\bj}$ is defined by $\Phi^{\bi}{}_{\bk}\Phi_{\bj}{}^{\bk}
=\delta^{\bi}_{\bj}$, then $e^\mu_{\bi}={}_0e^\mu_{\bj}\Phi^{\bj}{}_{\bi}$. 
$\Phi_{\bi}{}^{\bj}$ can be decomposed in a unique way as $\Phi_{\bi}{}^{\bj}
=S_{\bi}{}^{\bk}\Lambda_{\bk}{}^{\bj}$, where $\Lambda_{\bi}{}^{\bj}$ is a 
rotation matrix, $\Lambda_{\bi}{}^{\bk}\Lambda_{\bj}{}^{\bl}\eta_{\bk\bl}=
\eta_{\bi\bj}$, and $S_{\bi\bj}:=S_{\bi}{}^{\bk}\eta_{\bk\bj}=S_{(\bi\bj)}$. 
$\Lambda_{\bi}{}^{\bj}$ represents the pure 
gauge, while $s_{\bi}{}^{\bj}:=S_{\bi}{}^{\bj}-\delta_{\bi}^{\bj}$ the 
metric `deformation' content of $\Phi_{\bi}{}^{\bj}$. In fact, the `physical' 
metric is a quadratic expression of the symmetric part: 
$h_{\mu\nu}={}_0\zeta^{\bi}_\mu{}_0\zeta^{\bj}_\nu S_{\bi}{}^{\bk}S_{\bj}
{}^{\bl}\eta_{\bk\bl}={}_0h_{\mu\nu}+{}_0\zeta^{\bi}_{(\mu}{}_0\zeta^{\bj}
_{\nu)}(2s_{\bi\bj}+2s_{\bi}{}^{\bk}s_{\bj}{}^{\bl}\eta_{\bk\bl})$. The 
matrices in the decomposition $\Phi^{\bi}{}_{\bj}=S^{\bi}{}_{\bk}\Lambda
^{\bk}{}_{\bj}$ are transposed inverses of the corresponding matrices in 
$\Phi_{\bi}{}^{\bj}$: $\Lambda^{\bi}{}_{\bk}\Lambda_{\bj}{}^{\bk}=\delta
^{\bi}_{\bj}$ and $S^{\bi}{}_{\bk}S_{\bj}{}^{\bk}=\delta^{\bi}_{\bj}$. Note 
that although $\Lambda^{\bi}{}_{\bj}=\eta^{\bi\bk}\Lambda_{\bk}{}^{\bl}\eta
_{\bl\bj}$, $S^{\bi}{}_{\bj}$ is {\it not} $\eta^{\bi\bk}S_{\bk}{}^{\bl}\eta
_{\bl\bj}$. Then first calculate the connection coefficients $\Gamma^{\bi}
_{\mu{\bj}}$. They are 

$$
\Gamma^{\bi}_{\mu{\bj}}=-\bigl({}_0D_\mu\Phi_{\bk}{}^{\bi}\bigr)\Phi^{\bk}
{}_{\bj}+\Phi^{\bk}{}_{\bj}\Phi^{\bl}{}_{\bn}\eta^{\bn\bi}{}_0e^\rho_{\bk}
{}_0e^\nu_{\bl}{1\over2}\Bigl(-{}_0D_\nu h_{\mu\rho}+{}_0D_\mu h_{\rho\nu}
+{}_0D_\rho h_{\nu\mu}\Bigr). \eqno(2.2.1)
$$
\ni
Since we are interested in the fall-off of $h_{\mu\nu}$ and of $\chi_{\mu
\nu}$ implied by (2.1.2), we may write $S^{\bi}{}_{\bj}=\delta^{\bi}_{\bj}-
s_{\bj}{}^{\bi}$ and retain in (2.2.1) only the terms which are zeroth and 
first order in $s_{\bi\bj}$. We get $({}_0\zeta^{\bi}_\mu\Lambda_{\bi}{}
^{\bk})({}_0\zeta^{\bj}_\nu\Lambda_{\bj}{}^{\bl})\eta_{\bk\bn}\Gamma^{\bn}
_{\omega{\bl}}=-({}_0\zeta^{\bi}_\mu\Lambda_{\bi}{}^{\bk}){}_0\zeta^{\bj}
_\nu$ $(\partial_\omega\Lambda_{\bj\bk})+{}_0\zeta^{\bk}_\omega((\partial
_\mu s_{\bk\bl}){}_0\zeta^{\bl}_\nu-(\partial_\nu s_{\bk\bl}){}_0\zeta^{\bl}
_\mu)$. The evaluation of this 
equation for the various components yields the following results: First, the 
fact that the connection coefficients $\Gamma^{\bi}_{\mu{\bj}}$ come from 
a metric links the powers `{\rm a}' and `{\rm b}' in (2.1.2): a=b+1. Second, 
the fall-off rate of the metric is just that of the tangential components 
of the connection: $s_{\bi\bj}(r,\theta,\phi)=s_{\bi\bj}(\theta,\phi)/
r^{\rm b} +o(r^{-{\rm b}})$. Third, the gauge part must also tend to a 
constant gauge transformation with the same fall-off: $\Lambda_{\bi}{}^{\bj}
(r,\theta,\phi)={}_0\Lambda_{\bi}{}^{\bj}+\Lambda_{\bi}{}^{\bj}(\theta,\phi)
/r^{\rm b}+o(r^{-{\rm b}})$. Finally, taking into account these results, the 
equation $\chi_{\mu\nu}=-\Gamma^0_{\mu{\bj}}\zeta^{\bj}_\nu$ yields the 
fall-off $e^\mu_{\bi}e^\nu_{\bj}\chi_{\mu\nu}(r,\theta,\phi)=\chi_{\bi\bj}
(\theta,\phi)/r^{1+{\rm b}}+o(r^{-1-{\rm b}})$. Thus the Sen--Chern--Simons 
functional is well 
defined for those asymptotically flat initial data sets $(\Sigma,h_{\mu\nu},
\chi_{\mu\nu})$ for which both $h_{\mu\nu}-{}_0h_{\mu\nu}$ and $r\chi_{\mu
\nu}$ fall off like $r^{-{\rm b}}$ for some positive b. But since b may be 
arbitrarily small, the weakest possible fall-off for $h_{\mu\nu}-{}_0h_{\mu
\nu}$ and $r\chi_{\mu\nu}$ is logarithmic. \par
       The variational derivative of $Y_0$ with respect to $h_{\mu\nu}$ and 
$\chi_{\mu\nu}$ has been calculated [2]: 

$$
{\delta Y_0\over\delta\chi_{\mu\nu}}=8\sqrt{\vert h\vert}H^{\mu\nu}, 
\hskip 20pt
{\delta Y_0\over\delta h_{\mu\nu}}=-4\sqrt{\vert h\vert}\Bigl(B^{\mu\nu}+
\chi^{\rho(\mu}H^{\nu)}{}_\rho\Bigr), \eqno(2.2.2)
$$
\ni
where $H_{\mu\nu}$ is the conform magnetic curvature introduced above, and 
$B_{\mu\nu}:=-\varepsilon_{\rho\omega(\mu}D^\rho(R^\omega{}_{\nu)}+V^\omega
{}_{\nu)})+{1\over2}\chi^\rho{}_{(\mu}\varepsilon_{\nu)\rho\omega}(D_\lambda
\chi^{\omega\lambda}-D^\omega\chi)$. Both $H_{\mu\nu}$ and $B_{\mu\nu}$ are 
symmetric and trace free, and $B_{\mu\nu}$ reduces to the Cotton--York 
tensor of $(\Sigma,h_{\mu\nu})$ if $\chi_{\mu\nu}$ is vanishing. We have 
shown that the stationary points of $Y_0[h_{\mu\nu},\chi_{\mu\nu}]$ (i.e. 
for which $B_{\mu\nu}=0$ and $H_{\mu\nu}=0$) are precisely the data sets 
$(\Sigma,h_{\mu\nu},\chi_{\mu\nu})$ that can be locally isometrically 
imbedded into a conformally flat spacetime with first and second fundamental 
forms $h_{\mu\nu}$ and $\chi_{\mu\nu}$, respectively. \par
       The spacetime conformal rescaling of $(\Sigma,h_{\mu\nu},\chi_{\mu
\nu})$ by the pair of functions $\Omega:\Sigma\rightarrow(0,\infty)$, $\dot
\Omega:\Sigma\rightarrow{\bf R}$ was defined by the data set $(\Sigma,\hat
h_{\mu\nu},\hat\chi_{\mu\nu})$, where $\hat h_{\mu\nu}:=\Omega^2h_{\mu\nu}$ 
and $\hat\chi_{\mu\nu}:=\Omega\chi_{\mu\nu}+\dot\Omega h_{\mu\nu}$. To 
preserve the fall-off properties of $h_{\mu\nu}$ and $\chi_{\mu\nu}$ we 
should impose $\lim_{r\rightarrow\infty}\Omega=1$ and $\lim_{r\rightarrow
\infty}\dot\Omega=0$. In [2] we calculated the transformation of the 
connection coefficients $\Gamma^{\ua}_{\mu{\ub}}$, the curvature $F^a{}
_{b\mu\nu}$ and the Sen--Chern--Simons functional under spacetime conformal 
rescalings. If we define $\Upsilon_a:=\Omega^{-1}(\dot\Omega t_a+\Theta^\mu
_aD_\mu\Omega)$, then in the present notations $Y[\Gamma^{\ua}{}_{\ub}]$ 
transforms as 

$$
Y[\hat\Gamma^{\ua}{}_{\ub}]-Y[\Gamma^{\ua}{}_{\ub}]=-\int_\Sigma D_\rho
\Bigl(\varepsilon^{\rho\mu\nu}\Upsilon_aE^a_{\ua}\Gamma^{\ua}_{\mu{\ub}}
\vartheta^{\ub}_\nu\Bigr){\rm d}\Sigma,\eqno(2.2.3)
$$
\ni
where ${\rm d}\Sigma$ is the metric volume element on $\Sigma$. Thus $Y_0[h
_{\mu\nu},\chi_{\mu\nu}]$ is invariant with respect to the allowed spacetime 
conformal rescalings of the initial data sets. (From the right hand side of 
the corresponding formula in [2] the minus sign is missing.) Under the 
spacetime conformal rescaling the tensors $H_{\mu\nu}$ and $B_{\mu\nu}$ 
transform as $\hat H_{\mu\nu}=H_{\mu\nu}$ and $\hat B_{\mu\nu}=\Omega^{-1}
(B_{\mu\nu}+\Omega^{-1}\dot\Omega H_{\mu\nu})$. \par
      Finally, the diffeomorphism invariance of $Y_0$ implies that for any 
smooth vector field $N^\mu$, generating 1-parameter families of 
diffeomorphisms which preserve the fall-off properties of $h_{\mu\nu}$ and 
$\chi_{\mu\nu}$, the integral $\int_\Sigma({\delta Y_0\over\delta h_{\mu\nu}}
\L_{\bf N}h_{\mu\nu}+{\delta Y_0\over\delta\chi_{\mu\nu}}\L_{\bf N}\chi_{\mu
\nu}){\rm d}\Sigma$ is vanishing, where $\L_{\bf N}$ denotes the Lie 
derivative along $N^\mu$. This yields a divergence identity for $H_{\mu\nu}$ 
and $B_{\mu\nu}$. \par
\bigskip

\ni
{\bf 2.3 The spinor representation and the Ashtekar--Chern--Simons 
functional}\par
\bigskip
\ni
Let $\pi:S\rightarrow\Sigma$ be a trivializable principal fibre bundle over 
$\Sigma$ with structure group $SL(2,{\bf C})$, $\rho$ its defining 
representation on the two complex dimensional vector space ${\bf S}^A$, and 
$S^A(\Sigma)$ the (trivializable) associated vector bundle. Because of its 
trivializability there always exist globally defined frame fields $\{
\varepsilon^A_{\uA}\}$, $\uA=0,1$. Thus the capital Roman indices are 
abstract spinor indices, while the underlined capital Roman indices are 
name indices referring to a spinor basis. The dual, complex conjugate and 
the dual-complex conjugate bundles will be denoted by $S_A(\Sigma)$, $\bar 
S^{A'}(\Sigma)$ and $\bar S_{A'}(\Sigma)$, respectively, in which the 
global frame fields are $\{\varepsilon^{\uA}_A\}$, $\{\bar\varepsilon^{A'}
_{{\uA}'}\}$ and $\{\bar\varepsilon^{{\uA}'}_{A'}\}$. If $\epsilon_{{\uA}
{\uB}}$ is the alternating Levi-Civita symbol, then $\varepsilon_{AB}:=
\varepsilon^{\uA}_A\varepsilon^{\uB}_B\epsilon_{{\uA}{\uB}}$ defines a 
symplectic fibre metric on $S^A(\Sigma)$, i.e. with respect to which $\{
\varepsilon^A_{\uA}\}$ is normalized (or spin frame). By means of 
$\varepsilon_{AB}$ and its inverse $\varepsilon^{AB}$, defined by 
$\varepsilon^{AC}\varepsilon_{BC}=\delta^A_B$, $S^A(\Sigma)$ can be 
identified with $S_A(\Sigma)$, and, by the complex conjugate metric 
$\varepsilon_{A'B'}$ and its inverse $\varepsilon^{A'B'}$, $\bar S^{A'}
(\Sigma)$ can be identified with $\bar S_{A'}(\Sigma)$. \par
      Because of the trivializability of the bundles $V^a(\Sigma)$ and $S^A(
\Sigma)$ the well known isomorphism of the Lorentzian vector space and the 
space of Hermitian spinors, explained e.g. in [9], can be `globalized' on 
the whole of $\Sigma$. Namely, there exists a bundle isomorphism between 
the complexified Lorentzian vector bundle and the tensor product of $S^A
(\Sigma)$ with its complex conjugate bundle, $\vartheta:V^a(\Sigma)\otimes
{\bf C}\rightarrow S^A(\Sigma)\otimes\bar S^{A'}(\Sigma):$ $(p,X^a)\mapsto
(p,X^a\vartheta^{AA'}_a)$, satisfying $(\vartheta^{AA'}_a\vartheta^{BB'}_b
+\vartheta^{AA'}_b\vartheta^{BB'}_a)\varepsilon_{A'B'}=g_{ab}\varepsilon
^{AB}$. Therefore $\vartheta^{AA'}_a$ links the fibre metrics $\varepsilon
_{AB}$ on $S^A(\Sigma)$ and $g_{ab}$ on $V^a(\Sigma)$ and defines an $SL(2,
{\bf C})$-spinor structure on $\Sigma$ in the sense that $\pi:S\rightarrow
\Sigma$ is the universal covering bundle of $\pi:L\rightarrow\Sigma$ and 
$\vartheta^{AA'}_a$ maps the right action of $SL(2,{\bf C})$ on the former 
to the right action of $SO_0(1,3)$ on the latter. $S^A(\Sigma)$ is the 
bundle of $SL(2,{\bf C})$ spinors on $\Sigma$. The 
image of $V^a(\Sigma)$ in $S^A(\Sigma)\otimes\bar S^{A'}(\Sigma)$ is the 
bundle of Hermitian spinors. Thus, if $E^a_{AA'}$ is the inverse of 
$\vartheta^{AA'}_a$, then $E^a_{AA'}\lambda^A\bar\lambda^{A'}$ is real, 
null (with respect to the Lorentzian metric) and is either future or past 
directed for any spinor $\lambda^A$. We choose $\vartheta^{AA'}_a$ such 
that $E^a_{AA'}\lambda^A\bar\lambda^{A'}$ be {\it future} directed. 
Therefore the normal section $t_a$ of $V_a(\Sigma)$ defines a {\it positive} 
definite Hermitean fibre metric on $S^A(\Sigma)$ by $G_{AA'}:=\sqrt2 
t_{AA'}:=\sqrt2 t_aE^a_{AA'}$. As a consequence of the normalization $G_{A
A'}$ is compatible with the symplectic metric: $\varepsilon^{A'B'}G_{AA'}
G_{BB'}=\varepsilon_{AB}$, and hence $\varepsilon^{AB}\varepsilon^{A'B'}G
_{BB'}$ is just the inverse $G^{AA'}$ of $G_{AA'}$. If $\{\varepsilon^A
_{\uA}\}$ is a spin frame field and $\{\bar\varepsilon^{A'}_{{\uA}'}\}$ 
its complex conjugate then $E^a_{\ua}:=E^a_{AA'}\varepsilon^A_{\uA}\bar
\varepsilon^{A'}_{{\uA}'}\sigma^{{\uA}{\uA}'}_{\ua}$ is a $g
_{ab}$-orthonormal global frame field in $V^a(\Sigma)$, where $\sigma_{{\uA}
{\uB}'}^{\ua}$ are the $SL(2,{\bf C})$ Pauli matrices. \par
       Since any connection on a principal bundle determines a unique 
connection on each of its covering bundles (see e.g. Theorem 6.2 of Ch. II. 
in [10], which can be generalized to cover this case), the connection on 
$\pi:L\rightarrow\Sigma$ defines a connection on $\pi:S\rightarrow\Sigma$. 
Thus ${\cal D}_\mu$ on $V^a(\Sigma)$ determines a covariant derivative, 
denoted also by ${\cal D}_\mu$, both on $S^A(\Sigma)$ and $\bar S^{A'}
(\Sigma)$, and ${\cal D}_\mu$ annihilates both $\varepsilon_{AB}$ and 
$\varepsilon_{A'B'}$. This spinor covariant derivative is fixed 
completely by the requirement $(({\cal D}_\mu\lambda^A)\bar\mu^{A'}+\lambda
^A({\cal D}_\mu\bar\mu^{A'}))E^a_{AA'}={\cal D}_\mu(\lambda^A\bar\mu^{A'}
E^a_{AA'})$. Applying this equation to the spinors of the spin frame 
field we get the connection between the spinor connection coefficients 
$\Gamma^{\uA}_{\mu{\uB}}:=\varepsilon^{\uA}_A{\cal D}_\mu\varepsilon^A
_{\uB}$ and the Lorentzian connection coefficients: $\delta^{{\uA}'}
_{{\uB}'}\Gamma^{\uA}
_{\mu{\uB}}+\delta^{\uA}_{\uB}\bar\Gamma^{{\uA}'}_{\mu{\uB}'}=\sigma^{{\uA}
{\uA}'}_{\ua}\sigma_{{\uB}{\uB}'}^{\ub}\Gamma^{\ua}_{\mu{\ub}}$. Then for 
the curvature $F^A{}_{B\mu\nu}$ of ${\cal D}_\mu$ on $S^A(\Sigma)$, as can 
be expected, $\delta^{A'}_{B'}F^A{}_{B\mu\nu}+\delta^A_B\bar F^{A'}{}_{B'\mu
\nu}=\vartheta^{AA'}_aE^b_{BB'}F^a{}_{b\mu\nu}$ holds. Now we are in the 
position to be able to form the Chern--Simons 3-form and functional $Y[
\Gamma^{\uA}{}_{\uB}]$ built from the $SL(2,{\bf C})$-connection $\Gamma
^{\uA}_{\mu{\uB}}$ on $S^A(\Sigma)$, for which we get the following simple 
result: 

$$
Y[\Gamma^{\ua}{}_{\ub}]=2\Bigl(Y[\Gamma^{\uA}{}_{\uB}]+Y[\bar\Gamma^{{\uA}'}
{}_{{\uB}'}]\Bigr)=2\Bigl(Y[\Gamma^{\uA}{}_{\uB}]+\overline{Y[\Gamma^{\uA}
{}_{\uB}]}\Bigr). \eqno(2.3.1)
$$
\ni
We note that the fall-off conditions for the initial data determined in 
the previous subsection imply the existence of the functional $Y[\Gamma
^{\uA}{}_{\uB}]$, too. Thus the Sen--Chern--Simons functional is only the 
(four times the) real part of the Chern--Simons functional built from the 
$SL(2,{\bf C})$ spinor connection of the initial data set on $S^A(\Sigma)$. 
\par
      In [2] we also considered the self-dual/anti-self-dual representation 
$\rho_\pm$ of $SO_0(1,3)$ and the Chern--Simons functional on the 
corresponding vector bundle. That vector bundle was ${}^\pm\Lambda^2
(\Sigma)$, the bundle of self-dual/anti-self-dual 2-forms. $E^a_{AA'}$ 
defines an isomorphism between ${}^-\Lambda^2(\Sigma)$ and the bundle 
$S_{(AB)}(\Sigma)$ of the symmetric second rank unprimed spinors, and 
between ${}^+\Lambda^2(\Sigma)$ and $\bar S_{(A'B')}(\Sigma)$. If $\{
\varepsilon^{\uA}_A\}$ is a spin frame field in $S_A(\Sigma)$ then 
$\varepsilon^{\bi}_{AB}:=\sigma^{\bi}_{{\uA}{\uB}}\varepsilon^{\uA}_A
\varepsilon^{\uB}_B$, ${\bi}=1,2,3$, form a global frame field in $S_{(AB)}
(\Sigma)$, where $\sigma^{\bi}_{{\uA}{\uB}}$ are the $SU(2)$ Pauli matrices. 
This basis is orthonormal with respect to the scalar product $\langle z,w
\rangle=z_{AB}w_{CD}\varepsilon^{AC}\varepsilon^{BD}$ on $S_{(AB)}(\Sigma)$, 
inherited from the scalar product $\langle Z,W\rangle:={1\over2}Z_{ab}W_{cd}
g^{ac}g^{bd}$ on $\Lambda^2(\Sigma)$. (The frame field for and the scalar 
product on ${}^-\Lambda^2(\Sigma)$ that we used in [2] was $\sqrt2$ times 
and four times the `natural' choice $\varepsilon^{\bi}_{AB}\varepsilon_{A'
B'}\vartheta^{AA'}_a\vartheta^{BB'}_b$ and scalar product above, 
respectively.) The derivative ${\cal D}_\mu$ can naturally be extended 
to these bundles, whose connection coefficients in the basis $\{\varepsilon
^{\bi}_{AB}\}$ are ${}^-A^{\bi}_{\mu{\bj}}:=\varepsilon^{\bi}_{AB}{\cal D}
_\mu\varepsilon^{AB}_{\bj}=-{\rm i}\sqrt2 \varepsilon^{\bi}{}_{\bj\bk}\sigma
^{\bk}_{{\uA}{\uB}}\Gamma^{{\uA}{\uB}}_\mu$, and similarly in the complex 
conjugate basis $\{\bar\varepsilon^{\bi}_{A'B'}\}$ they are ${}^+A^{\bi}
_{\mu{\bj}}:=\bar\varepsilon^{\bi}_{A'B'}{\cal D}_\mu\bar\varepsilon^{A'B'}
_{\bj}={\rm i}\sqrt2 \varepsilon^{\bi}{}_{\bj\bk}\bar\sigma^{\bk}_{{\uA}'
{\uB}'}\bar\Gamma^{{\uA}'{\uB}'}_\mu=\overline{{}^-A^{\bi}_{\mu{\bj}}}$. 
Then the corresponding curvature 2-forms are ${}^-F^{\bi}{}_{{\bj}\mu\nu}=
-{\rm i}\sqrt2 \varepsilon^{\bi}{}_{\bj\bk}\varepsilon^{\bk}_{AB}F^{AB}
{}_{\mu\nu}$, and ${}^+F^{\bi}{}_{{\bj}\mu\nu}=\overline{{}^-F^{\bi}
{}_{{\bj}\mu\nu}}$. The Chern--Simons functional built from ${}^\pm A^{\bi}
_{\mu{\bj}}$, which we called in [2] the self-dual/anti-self-dual 
Ashtekar--Chern--Simons functional, can now be reexpressed by the $SL(2,
{\bf C})$ connection and curvature: 

$$
Y[{}^-A^{\bi}{}_{\bj}]=4Y[\Gamma^{\uA}{}_{\uB}], \hskip 20pt
Y[{}^+A^{\bi}{}_{\bj}]=4Y[\bar\Gamma^{{\uA}'}{}_{{\uB}'}]=
 4\overline{Y[\Gamma^{\uA}{}_{\uB}]}. \eqno(2.3.2)
$$
\ni
Thus, as could be expected, the Ashtekar--Chern--Simons functional is 
essentially the Chern--Simons functional built from the $SL(2,{\bf 
C})$-spinor connection, and the Sen--Chern--Simons functional is just its 
real part. The functional derivative of $Y_\pm:=Y_\pm[h_{\mu\nu},\chi_{\mu
\nu}]$, defined by $Y[{}^\pm A^{\bi}{}_{\bj}]$ modulo $16\pi^2$, with 
respect to $\chi_{\mu\nu}$ and $h_{\mu\nu}$ are 

$$
\eqalign{
{\delta Y_\pm\over\delta\chi_{\mu\nu}}=&8\sqrt{\vert h\vert}\Bigl(H^{\mu\nu}
  \mp{\rm i}\bigl(R^{\mu\nu}-{1\over2}Rh^{\mu\nu}+V^{\mu\nu}-{1\over2}V
  h^{\mu\nu}\bigr)\Bigr), \cr
{\delta Y_\pm\over\delta h_{\mu\nu}}=&-4\sqrt{\vert h\vert}\bigl(B^{\mu\nu}
  +\chi^{(\mu}{}_\rho H^{\nu)\rho}\bigr)\pm 4{\rm i}\sqrt{\vert h\vert}
  \Bigl(\varepsilon^{\rho\omega(\mu}D_\rho H^{\nu)}{}_\omega+{1\over2}D^{
  (\mu}\bigl(D_\rho\chi^{\nu)\rho}-D^{\nu)}\chi\bigr)-\cr
&-{1\over2}h^{\mu\nu}D_\omega\bigl(D_\rho\chi^{\rho\omega}-D^\omega\chi
 \bigr)+\chi^{(\mu}{}_\rho\bigl(R^{\nu)\rho}-{1\over2}h^{\nu)\rho}R+V^{\nu)
  \rho}-{1\over2}h^{\nu)\rho}V\bigr)\Bigr).\cr} \eqno(2.3.3)
$$
\ni
The stationary points of these functionals were shown to be those data sets 
that can be locally isometrically imbedded into the Minkowski spacetime with 
first and second fundamental forms $h_{\mu\nu}$ and $\chi_{\mu\nu}$, 
respectively. Under spacetime conformal rescalings $Y_\pm$ are {\it not} 
invariant, because e.g. under the infinitesimal conformal rescaling 
$\delta h_{\mu\nu}=2h_{\mu\nu}\delta\Omega$, $\delta\chi_{\mu\nu}=\chi_{\mu
\nu}\delta\Omega+h_{\mu\nu}\delta\dot\Omega$, they transform as $\delta 
Y_\pm=\pm8{\rm i}\int_\Sigma({1\over2}(R+V)\delta\dot\Omega-D_\mu(D_\nu
\chi^{\mu\nu}-D^\mu\chi)\delta\Omega){\rm d}\Sigma$. (Unfortunately the 
formulae corresponding to (2.3.3) in [2] contain trivial numerical and 
sign errors, and consequently the expression for $\delta Y_\pm$ given 
there is also erroneous.) \par
     Next, let us consider general finite dimensional {\it irreducible} 
representations of $SL(2,{\bf C})$, the corresponding associated vector 
bundles, the connection on them, and the Chern-Simons functional. It is 
well known that any finite dimensional 
irreducible representation of $SL(2,{\bf C})$ is characterized by a pair 
$(k,l)$ of nonnegative integers and the representation space is ${\bf S}^{
(A_1...A_k)}\otimes\bar{\bf S}^{(B'_1...B'_l)}$, the tensor product of the 
space of the totally symmetric spinors of rank $k$ and of the totally 
symmetric primed spinors of rank $l$. Thus a basis in this space has the 
form $\varepsilon_{\bi}^{A_1...A_kB'_1...B'
_l}=\varepsilon_{\bi}^{(A_1...A_k)(B'_1...B'_l)}$, ${\bi}=1,...,(k+1)(l+1)$. 
The dual basis in the dual space ${\bf S}_{(A_1...A_k)}\otimes\bar{\bf S}
_{(B'_1...B'_l)}$ is denoted by $\varepsilon^{\bi}_{A_1...A_kB'_1...B'_l}$. 
These bases can also be expressed by the tensor product bases: 
$\varepsilon_{\bi}^{A_1...A_kB'_1...B'_l}=\sigma_{\bi}^{{\uA}_1...{\uA}_k
{\uB}'_1...{\uB}'_l}\varepsilon^{A_1}_{{\uA}_1}...\varepsilon^{A_k}_{{\uA}
_k}\bar\varepsilon^{B'_1}_{{\uB}'_1}...\bar\varepsilon^{B'_l}_{{\uB}'_l}$ 
and $\varepsilon^{\bi}_{A_1...A_kB'_1...B'_l}=\sigma^{\bi}_{{\uA}_1...{\uA}
_k{\uB}'_1...{\uB}'_l}\varepsilon_{A_1}^{{\uA}_1}...\varepsilon_{A_k}^{
{\uA}_k}\bar\varepsilon_{B'_1}^{{\uB}'_1}...\bar\varepsilon_{B'_l}^{{\uB}'
_l}$, where the combination coefficients (the well known Clebsch--Gordan 
coefficients) are completely symmetric both in their unprimed and primed 
spinor name indices 
and satisfy the duality conditions $\sigma^{\bi}_{{\uA}_1...{\uA}_k{\uB}'_1
...{\uB}'_l}\sigma_{\bj}^{{\uA}_1...{\uA}_k{\uB}'_1...{\uB}'_l}=\delta^{\bi}
_{\bj}$ and $\sigma^{\bi}_{{\uA}_1...{\uA}_k{\uB}'_1...{\uB}'_l}\sigma_{\bi}
^{{\uC}_1...{\uC}_k{\uD}'_1...{\uD}'_l}=\delta^{{\uC}_1}_{({\uA}_1}...\delta
^{{\uC}_k}_{{\uA}_k)}\delta^{{\uD}'_1}_{({\uB}'_1}...\delta^{{\uD}'_l}_{
{\uB}'_l)}$. Then the connection coefficients of the connection on the 
associated vector bundle $S^{(A_1...A_k)(B'_1...B'_l)}(\Sigma)$, determined 
by the connection ${\cal D}_\mu$ on $S^A(\Sigma)$, are 

$$\eqalign{
A^{\bi}_{\mu{\bj}}:&=\varepsilon^{\bi}_{A_1...A_kB'_1...B'_l}
  {\cal D}_\mu\varepsilon^{A_1...A_kB'_1...B'_l}_{\bj}=\cr
&=k\sigma^{\bi}_{{\uE}{\uA}_2...{\uA}_k{\uB}'_1...{\uB}'_l}\sigma_{\bj}
  ^{{\uF}{\uA}_2...{\uA}_k{\uB}'_1...{\uB}'_l}\Gamma^{\uE}_{\mu{\uF}}+
  l\sigma^{\bi}_{{\uA}_1...{\uA}_k{\uE}'{\uB}'_2...{\uB}'_l}\sigma_{\bj}
  ^{{\uA}_1...{\uA}_k{\uF}'{\uB}'_2...{\uB}'_l}\bar\Gamma^{{\uE}'}_{\mu
  {\uF}'},\cr}\eqno(2.3.4)
$$
\ni
and there is a similar relation between the curvature 2-forms $F^{\bi}{}
_{{\bj}\mu\nu}$ and $F^{\uE}{}_{{\uF}\mu\nu}$. Then, using the duality 
conditions for the Clebsch--Gordan coefficients, the Chern--Simons 
functional $Y[A^{\bi}{}_{\bj}]$ defined on the vector bundle $S^{
(A_1...A_k)(B'_1...B'_l)}(\Sigma)$ can be computed easily: 

$$
Y[A^{\bi}{}_{\bj}]={1\over6}(k+1)(l+1)\Bigl(k(k+2)Y[\Gamma^{\uA}{}
_{\uB}]+l(l+2)\overline{Y[\Gamma^{\uA}{}_{\uB}]}\Bigr).\eqno(2.3.5)
$$
\ni
Thus this is a combination of $Y[\Gamma^{\uA}{}_{\uB}]$ and its complex 
conjugate with integers depending on the representation. The 
representations in which the Chern--Simons functional is conformally 
invariant are precisely the tensor representations; i.e. for which $l=k$. 
Therefore the higher order irreducible representations do not give anything 
new. The Sen--Chern--Simons and the anti-self-dual/self-dual 
Ashtekar--Chern--Simons functionals correspond to the (1,1), (2,0) and 
(0,2) cases, and by an appropriate choice for the normalization the 
Clebsch--Gordan coefficients reduce to the $SL(2,{\bf C})$ Pauli matrices, 
the $SU(2)$ Pauli matrices and their complex conjugate, respectively. \par
     Finally, since $SL(2,{\bf C})$ is semisimple, any finite dimensional 
representation $\rho$ of that is the direct sum of 
irreducible representations $\rho_1$,...,$\rho_n$ (i.e. it is completely 
reducible). But then the associated vector bundle $E(\Sigma)$ defined by 
$\rho$ is the Whitney sum of the vector bundles corresponding to 
$\rho_1$,...,$\rho_n$, and the connection on $E(\Sigma)$ is the sum of the 
connections of the constituent bundles. Hence the Chern--Simons functional 
defined on $E(\Sigma)$ is the sum of the Chern--Simons functionals of the 
constituent bundles. \par
      As is usual in the recent spinor approaches in general relativity in 
the rest of this paper we will not write out the isomorphisms $E^a_{AA'}$, 
$\vartheta^{AA'}_a$ and the Pauli matrices explicitly. Any Lorentzian 
tensor index, e.g. $a$ and ${\ua}$, can be freely replaced by the 
corresponding pair of spinor indices, i.e. by $AA'$ and ${\uA}{\uA}'$, 
respectively. \par

\bigskip
\ni
{\lbf 3. Relation to 3-surface twistors}\par
\bigskip

\ni
{\bf 3.1 The unitary form of ${\cal D}_\mu$ and the 3-surface twistors}\par
\bigskip
\ni
The positive definite Hermitean metric defines the bundle maps $\bar S_{A'}
(\Sigma)\rightarrow S_X(\Sigma):(p,\mu_{A'})\mapsto(p,G_X{}^{A'}\mu_{A'})$ 
and $\bar S^{A'}(\Sigma)\rightarrow S^X(\Sigma):(p,\mu^{A'})\mapsto(p,-G^X
{}_{A'}\mu^{A'})$. They are {\bf C}-linear isomorphisms taking the 
symplectic fibre metrics into the symplectic fibre metrics, and making 
possible to use only the unprimed spinors. In particular, the complex 
conjugation is represented by the {\bf C}-anti-linear operation ${}^\dagger:
S_A(\Sigma)\rightarrow S_A(\Sigma): (p,\lambda_A)
\mapsto(p,\lambda^\dagger_A):=(p,G_A{}^{A'}\bar\lambda_{A'})$, whose action 
can obviously be extended to arbitrary spinors. If we convert the primed 
name indices into unprimed ones in an analogous way, then these bundle maps 
take the dual complex conjugate and complex conjugate frames, $\{\bar
\varepsilon^{{\uA}'}_{A'}\}$ and $\{\bar\varepsilon^{A'}_{{\uA}'}\}$, into 
the frames $\{\varepsilon^{\uX}_X\}$ and $\{\varepsilon^X_{\uX}\}$, 
respectively. Every Lorentzian index corresponds to a pair of unprimed 
spinor indices, e.g. $X^a$ to $X^{AX}$; and $X^a$ is proportional to the 
normal section iff $X^{AX}=X^{[AX]}$, and $X^a=P^a_bX^b$ iff $X^{AB}=
X^{(AX)}$. Therefore the order of the unitary spinor indices, e.g. $AX$ 
above, is important unless the corresponding Lorentzian index is purely 
spatial. For example the unitary spinor form of the normal section 
$t_a$ and of the imbedding $\Theta^a_\mu$ are $t_{AX}:=G_X{}^{A'}t_{AA'}={1
\over\sqrt2}\varepsilon_{AX}$ and $\Theta^{AX}_\mu:=-\Theta^{AA'}_\mu G_{A'}
{}^X=\Theta^{(AX)}_\mu$, respectively. The unitary spinor form of other 
important tensor fields, namely the metric, the corresponding volume form 
and the extrinsic curvature are $h_{AXBY}=-\varepsilon_{A(B}\varepsilon
_{Y)X}$, $\varepsilon_{AXBYCZ}={{\rm i}\over\sqrt2}(\varepsilon_{A(B}
\varepsilon_{Y)(C}\varepsilon_{Z)X}+\varepsilon_{X(B}\varepsilon_{Y)(C}
\varepsilon_{Z)A})$ and $\chi_{\mu AX}:=G_X{}^{A'}\chi_{\mu AA'}=\chi_{\mu 
(AX)}$, respectively. A (2r,2s) type spinor $T^{A_1X_1...A_rX_r}_{B_1Y_1...
B_sY_s}$ is the unitary spinor form of a {\it real} Lorentz tensor iff 
$T^\dagger{}^{A_1X_1...A_rX_r}_{B_1Y_1...B_sY_s}=(-)^{r+s}T^{A_1X_1...A_r
X_r}_{B_1Y_1...B_sY_s}$. In particular, the reality of $\Theta^a_\mu$ 
implies $\Theta^{\dagger AB}_\mu=-\Theta^{AB}_\mu$. The square of the 
adjoint operation is $\xi^{\dagger\dagger}{}^{A_1...A_r}_{B_1...B_s}=(-)^{r
+s}\xi^{A_1...A_r}_{B_1...B_s}$. The action of the Sen derivative on the 
spinor fields can be written as ${\cal D}_\mu\lambda^A=D_\mu\lambda^A-{1
\over\sqrt2}\chi_\mu{}^A{}_B\lambda^B$. Note, however, that although both 
${\cal D}_\mu$ and $D_\mu$ annihilate $\varepsilon_{AB}$ and $D_\mu$ 
annihilates $G_{AA'}$, the Sen derivative doesn't annihilate the Hermitean 
metric: ${\cal D}_\mu G_{AA'}=\sqrt2\chi_{\mu AA'}\not=0$. Thus the 
operation of taking the unitary form of a spinor and the Sen--derivative 
are not commuting. Consequently ${\cal D}_\mu\lambda^{\dagger A}$ differs 
from $({\cal D}_\mu\lambda^A)^\dagger$, and hence it seems useful to 
introduce two Sen operators on the spinor fields, the self-dual and the 
anti-self-dual ones: ${}^\pm{\cal D}_\mu\lambda^A:=D_\mu\lambda^A\pm{1\over
\sqrt2}\chi_\mu{}^A{}_B\lambda^B$. Then $({}^\pm{\cal D}_\mu\lambda^{...}
_{...})^\dagger={}^\mp{\cal D}_\mu\lambda^\dagger{}^{...}_{...}$; i.e. 
they are adjoint to each other. (For a more detailed discussion of the 
theory of unitary spinors in relativity see e.g. [12-16].)\par
      One can define the unitary spinor form of the Sen operators too: 
${}^\pm{\cal D}_{AX}\lambda_B:=\Theta^\mu_{AX}{}^\pm{\cal D}_\mu\lambda_B=
D_{AX}\lambda_B\mp{1\over\sqrt2}\chi_{AXB}{}^C\lambda_C$, for which 
$({}^\pm{\cal D}_{AB}\lambda^{...}_{...})^\dagger=-{}^\mp{\cal D}_{AB}
\lambda^\dagger{}^{...}_{...}$. The commutator of two such operators is 

$$
{}^\pm{\cal D}_{X(A}{}^\pm{\cal D}^X_{B)}\lambda_C=-\lambda^D{}^\pm\Phi
_{DCAB}-{}^\pm\Psi^{EF}{}_{AB}{}^\pm{\cal D}_{EF}\lambda_C, \eqno(3.1.1)
$$
\ni
where 

$$\eqalignno{
{}^\pm\Psi^{CD}{}_{AB}&:=\mp{1\over\sqrt{2}}\Bigl(\chi^{CD}{}_{AB}-\chi
   \delta^C_{(A}\delta^D_{B)}\Bigr), &(3.1.2)\cr
{}^\pm\Phi_{ABCD}&:={1\over2}\Bigl(R_{ABCD}+{R\over2}\varepsilon_{A(C}
  \varepsilon_{D)B}\Bigr)\mp\cr
&\mp{1\over\sqrt2}D_{X(C}\chi^X{}_{D)AB}-{1\over4}\Bigl(\chi_C{}^{EF}{}_B
  \chi_{DEFA}+\chi_D{}^{EF}{}_B\chi_{CEFA}\Bigr), &(3.1.3)\cr}
$$
\ni
and $R_{ABCD}$ is the unitary spinor form of the Ricci tensor of $(\Sigma,
h_{\mu\nu})$. They represent the `torsion' and the curvature of ${}^\pm{\cal 
D}_\mu$, respectively. The latter is related to the curvature 2-form ${}^\pm
F^A{}_{B\mu\nu}$ of ${}^\pm{\cal D}_\mu$ by ${}^\pm\Phi_{ABCD}={1\over2}
{}^\pm F_{ABCRDS}\varepsilon^{RS}$. Their algebraic symmetries are 
${}^\pm\Psi_{ABCD}={}^\pm\Psi_{(AB)(CD)}={}^\pm\Psi_{CDAB}$ and ${}^\pm\Phi
_{ABCD}={}^\pm\Phi_{(AB)(CD)}$, and their contractions are 

$$\eqalignno{
{}^\pm\Psi_{ACB}{}^C&=\pm{1\over\sqrt2}\chi\varepsilon_{AB}, &(3.1.4)\cr
{}^\pm\Phi_{ACB}{}^C&=-\varepsilon_{AB}{1\over8}\Bigl(R+\chi^2-\chi_{\mu\nu}
   \chi^{\mu\nu}\Bigr)\pm{1\over2\sqrt2}\Bigl(D_\mu\chi^\mu{}_{AB}-D_{AB}
   \chi\Bigr).  &(3.1.5)\cr}
$$
\ni
Thus ${}^\pm\Phi_{ABCD}$ is {\it not} symmetric in the pairs $AB$ and $CD$. 
The unitary spinor form of the tensor fields $H_{\mu\nu}$ and $B_{\mu\nu}$ 
of subsection 2.2 is 

$$\eqalignno{
H_{AXBY}={{\rm i}\over\sqrt2}\Bigl(&D_{E(A}\chi^E{}_{X)BY}+D_{E(B}\chi^E
  {}_{Y)AX}\Bigr)={\rm i}\sqrt{2}D_{E(A}\chi^E{}_{XBY)}=\cr
={{\rm i}\over2}\Bigl(&{}^-\Phi_{AXBY}+{}^-\Phi_{BYAX}-{}^+\Phi_{AXBY}-{}^+
  \Phi_{BYAX}\Bigr), &(3.1.6) \cr
B_{AXBY}={{\rm i}\over2\sqrt2}\Bigl(&D_{EA}\bigl({}^\pm\Phi^E{}_{XBY}+{}^\pm
  \Phi_{BYX}{}^E\bigr)+D_{EX}\bigl({}^\pm\Phi^E{}_{ABY}+{}^\pm\Phi_{BYA}{}^E
  \bigr)+\cr
+&D_{EB}\bigl({}^\pm\Phi^E{}_{YAX}+{}^\pm\Phi_{AXY}{}^E\bigr)+D_{EY}\bigl(
  {}^\pm\Phi^E{}_{BAX}+{}^\pm\Phi_{AXB}{}^E\bigr)+\cr
+&\chi_{AX(B}{}^E\bigl(D^{CD}\chi_{Y)ECD}-D_{Y)E}\chi\bigr)+\chi_{BY(A}{}^E
  \bigl(D^{CD}\chi_{X)ECD}-D_{X)E}\chi\bigr)\Bigr)\mp\cr
\mp{1\over\sqrt2}\Bigl(&D^E_{(A}H_{X)EBY}+D^E_{(B}H_{Y)EAX}\Bigr). &(3.1.7) 
\cr}
$$
\ni
The second Bianchi identity for ${}^\pm F^A{}_{B\mu\nu}$ then takes the form 

$$
D_{AB}{}^\pm\Phi^{CDAB}\pm\sqrt2\chi^{(C}{}_{EAB}{}^\pm\Phi^{D)EAB}=0. 
\eqno(3.1.8)
$$
\ni
The first Bianchi identity doesn't give any further algebraic symmetry for 
${}^\pm\Phi_{ABCD}$. \par 
       Let $(M,g_{ab})$ be a Lorentzian spacetime manifold with spinor 
structure, and recall [9] that a contravariant 1-valence twistor field 
${\tt Z}^\alpha$ is a pair $(\omega^A,\pi_{A'})$ of spinor fields such 
that, under the conformal rescaling $\varepsilon_{AB}\mapsto\hat\varepsilon
_{AB}:=\Omega\varepsilon_{AB}$, the spinor fields transform as $\omega^A
\mapsto\hat\omega^A:=\omega^A$ (i.e. $\omega^A$ has zero conformal weight) 
and $\pi_{A'}\mapsto\hat\pi_{A'}:=\pi_{A'}+{\rm i}\omega^A\Upsilon_{AA'}$, 
where $\Upsilon_{AA'}:=\nabla_{AA'}\ln\Omega$. The spinor parts of ${\tt Z}
^\alpha$ will also be denoted by ${\tt Z}^A$ and ${\tt Z}_{A'}$. ${\tt Z}
^\alpha$ may be defined only on a submanifold of $M$, e.g. on a (say, 
spacelike) hypersurface $\theta(\Sigma)$ or on a spacelike 2-surface 
$\theta(\$)$. If $\omega^A$ is any spinor field on $M$ (or on $\theta(
\Sigma)$) with zero conformal weight, and if $\pi_{A'}$ is defined by ${1
\over2}{\rm i}\nabla_{A'A}\omega^A$ (or on $\theta(\Sigma)$ by ${2\over3}
{\rm i}{\cal D}_{A'A}\omega^A$, where ${\cal D}_a:=P^b_a\nabla_b$ is the 
three dimensional Sen connection [12]), then ${\tt Z}^\alpha:=(\omega^A,\pi
_{A'})$ turns out to be a twistor field on $M$ (or, respectively, on $\theta
(\Sigma)$) in the sense above. Thus any such spinor field $\omega^A$ on $M$ 
(or on $\Sigma$) determines a twistor field (`geometric twistor fields'). 
(If $\omega^A$ is a spinor field on $\theta(\$)$ with 
zero conformal weight and $\pi_{A'}:={\rm i}\Delta_{A'A}\omega^A$, where 
$\Delta_a:=\Pi^b_a\nabla_b$ is the projection of $\nabla_a$ to $\theta(\$)$, 
the two dimensional Sen connection [17], then ${\tt Z}^\alpha:=(\omega^A,
\pi_{A'})$ is a twistor field on $\theta(\$)$ in the sense above. This 
case, however, will not be considered in the present paper.) A geometric 
twistor field ${\tt Z}^\alpha$ on $M$ is called a global twistor, or simply 
twistor, if its primary spinor part $\omega^A$ is a solution of the 
1-valence twistor equation $\nabla^{A'(A}\omega^{B)}=0$. It is known [9] 
that if $\omega^A$ is a nonzero solution of the twistor equation, then it 
is a 4-fold principal spinor of the Weyl spinor and $\omega^A\bar\omega
^{A'}$ is a future pointing conformal Killing vector; and, conversely [18], 
if $\omega^A$ is a 4-fold principal spinor of $\psi_{ABCD}$ and $\omega^A
\bar\omega^{A'}$ is a future pointing conformal Killing vector then for 
some real function $f$ the spinor $\exp({\rm i}f)\omega^A$ is a solution 
of the twistor equation. Furthermore the twistor equation is completely 
integrable, i.e. admits the maximal number of solutions, namely four, if 
and only if $(M,g_{ab})$ is conformally flat. \par
       If ${\tt Z}^\alpha=(\omega^A,\pi_{A'})$ is any twistor field on the 
spacelike hypersurface $\theta(\Sigma)$, then its secondary part $\pi_{A'}$ 
can equivalently be represented by the unprimed spinor $\pi_X:=G_X{}^{A'}\pi
_{A'}$. Furthermore, by $G_Z{}^{A'}\nabla_{A'A}\omega_B={1\over\sqrt2}
\varepsilon_{AZ}(t^e\nabla_e\omega_B)+{}^-{\cal D}_{(ZA}\omega_{B)}-{2\over3}
\varepsilon_{B(A}{}^-{\cal D}_{Z)C}\omega^C$ the full 3+1 decomposition of 
the twistor equation is 

$$
{}^-{\cal D}_{(AB}\omega_{C)}=0, \hskip 20pt 
t^e\nabla_e\omega_A={\sqrt2\over3}{}^-{\cal D}_{AB}\omega^B. 
\eqno(3.1.9.a,b)
$$
\ni
The first of these, i.e. the spatial part of the twistor equation, is 
called the 3-surface twistor equation [6,19]. A geometric twistor field 
${\tt Z}^\alpha=(\omega^A,\pi_X)$ on $\theta(\Sigma)$ is called a (global) 
3-surface twistor if its primary spinor part $\omega^A$ is a solution of 
(3.1.9.a). As it was proved in [6], it is completely integrable, and hence 
admits four {\bf C}-linearly independent solutions, if and only if $\Sigma$ 
with its first and second fundamental forms can also be imbedded into a 
conformally flat spacetime. The self-dual Sen operator ${}^+{\cal D}_{AB}$ 
appears in the 3+1 decomposition of the complex conjugate twistor equation 
$\nabla^{A(A'}\omega^{B')}=0$. \par
\bigskip

\ni
{\bf 3.2 The 3-surface twistor connection and Chern--Simons functional}\par
\bigskip
\ni
The basic idea [9] of the twistor parallel transport on $M$ is to consider 
the global twistors as {\it constant} twistor fields with respect to the 
twistor connection. This idea was used to introduce the notion of 
3-surface twistor connection on (spacelike or timelike) hypersurfaces 
[6], by means of which the complete integrability of the 3-surface twistor 
equation could be characterized as the vanishing of the 3-surface twistor 
curvature. Since we need the explicit form both of the twistor connection 
and curvature we first recall the main points of the construction of the 
3-surface twistor connection in the unitary spinor formalism on an 
arbitrary triple $(\Sigma,h_{\mu\nu},\chi_{\mu\nu})$, then calculate the 
connection coefficients and the curvature, and finally we calculate the 
Chern--Simons 3-form built from this connection. \par
      The 3-surface twistor equation and its complex conjugate can be 
rewritten as 

$$
{}^\pm{\cal D}_{AB}\omega^C+{\rm i}\delta^C_{(A}\pi_{B)}=0. \eqno(3.2.1)
$$
\ni
Taking the ${}^\pm{\cal D}_{AB}$-derivative of its contraction $\pi_C={2
\over3}{\rm i}{}^\pm{\cal D}_{CD}\omega^D$, using the commutator (3.1.1) and 
the 3-surface twistor equation (3.2.1), and finally adding the resulting 
equation to itself after appropriate permutations of its indices we get 

$$
{}^\pm{\cal D}_{AB}\pi_C\pm\sqrt2\pi_D\chi^D{}_{CAB}-2{\rm i}\omega^D
{}^\pm\Phi_{D(AB)C}=0.   \eqno(3.2.2)
$$
\ni
The twistor field ${\tt Z}^\alpha=(\omega^A,\pi_X)$ on $\Sigma$ is a global 
3-surface twistor (or conjugate twistor according to the sign $\mp$) if and 
only if the spinor fields $\omega^A$ and $\pi_X$ satisfy the system of 
equations (3.2.1), (3.2.2). Thus, recalling that ${}^-{\cal D}_{AB}$ can be 
expressed by ${}^+{\cal D}_{AB}$ and the extrinsic curvature, it seems 
natural to define the covariant derivative of any twistor field ${\tt Z}
^\alpha$, defined on $\Sigma$, by the pair 

$$
{}^\pm D_{MN}{\tt Z}^\alpha:=
\Bigl({}^\pm{\cal D}_{MN}\omega^A+{\rm i}\delta^A_{(M}\pi_{N)},\,
{}^\mp{\cal D}_{MN}\pi_X-2{\rm i}\omega^B{}^\pm\Phi_{B(MN)X}\Bigl). 
\eqno(3.2.3)
$$
\ni
In fact, ${\tt Z}^\alpha$ is a global 3-surface twistor/conjugate twistor 
on $\Sigma$ iff $v^\mu{}^\mp D_\mu{\tt Z}^\alpha=0$ for any tangent vector 
$v^\mu$ of $\Sigma$, under spacetime conformal rescalings $v^\mu{}^\pm D
_\mu{\tt Z}^\alpha$ transform as twistor fields on $\Sigma$, and ${}^\pm 
D_\mu{\tt Z}^\alpha$ are determined by $(\Sigma,h_{\mu\nu},\chi_{\mu\nu})$. 
(If $\Sigma$ were imbedded in $(M,g_{ab})$ as a spacelike hypersurface 
then we could define ${\cal D}_a{\tt Z}^\alpha:=P^b_a\nabla_b{\tt Z}
^\alpha$, the projection to $\Sigma$ of the spacetime twistor covariant 
derivative as a derivative analogous to the Sen connection. That 
derivative, however, would depend not only on the data set $(\Sigma,h_{\mu
\nu},\chi_{\mu\nu})$ but on the spatial-spatial part of the spacetime Ricci 
tensor too.) The 3-surface twistor connection coefficients are defined by 
${}^\pm{\tt A}^{\ualpha}_{\mu{\ubeta}}:={\tt E}^{\ualpha}_\alpha{}^\pm D_\mu
{\tt E}^\alpha_{\ubeta}$ in the dual twistor frame fields $\{{\tt E}^\alpha
_{\ualpha}\}$, $\{{\tt E}^{\ubeta}_\beta\}$, which may be chosen to be 
determined by the spin frame fields $\{\varepsilon^A_{\uA}\}$, $\{
\varepsilon^{\uA}_A\}$ through ${\tt E}^\alpha_{\uA}=(\varepsilon^A_{\uA},0)$, ${\tt E}^\alpha{}
^{\uX}=(0,\varepsilon^{\uX}_X)$ and ${\tt E}^{\uB}_\beta=(\varepsilon^{\uB}
_B,0)$, ${\tt E}_\beta{}_{\uY}=(0,\varepsilon^Y_{\uY})$. They are 
represented by the following spinor parts 

$$
{}^\pm{\tt A}^{\ualpha}_{\mu{\ubeta}}=
\left(\matrix{{}^\pm{\tt A}_\mu{}^{\uA}{}_{\uB}& {}^\pm{\tt A}_\mu{}^{
   {\uA}{\uY}}\cr   
   {}^\pm{\tt A}_\mu{}_{{\uX}{\uB}}& {}^\pm{\tt A}_\mu{}_{\uX}{}^{\uY}
   \cr}\right)=\left(
\matrix{{}^\pm\Gamma^{\uA}_{\mu{\uB}}& {\rm i}\Theta^{{\uA}{\uY}}_\mu\cr
 -2{\rm i}\varepsilon^B_{\uB}\varepsilon^X_{\uX}{}^\pm\Phi_{B(MN)X}\Theta
 ^{MN}_\mu  & -{}^\mp\Gamma^{\uY}_{\mu{\uX}}\cr}\right). \eqno(3.2.4)
$$
\ni
Here ${}^\pm\Gamma^{\uA}_{\mu{\uB}}:=\varepsilon^{\uA}_A{}^\pm{\cal D}_\mu
\varepsilon^A_{\uB}$ are the connection coefficients of ${}^\pm{\cal D}
_\mu$ in the dual spin frame fields. Next, using formulae 
(3.1.2)-(3.1.8) and (3.2.3)-(3.2.4), a rather laborious calculation yields 
the 3-surface twistor curvature. The only non-vanishing spinor parts of 
that curvature are 

$$\eqalignno{
{}^\pm{\tt R}_X{}^Y{}_{\mu\nu}=&\pm2{\rm i}H^Y{}_{XMN}\varepsilon_{RS}
  \Theta^{MR}_\mu\Theta^{NS}_\nu,   &(3.2.5)\cr
{}^\pm{\tt R}_{XB\mu\nu}=&\Bigl(\sqrt2 B_{XBMN}\pm\bigl(D^E_{(X}H_{B)EMN}+
  D^E_{(M}H_{N)EXB}\bigr)+\cr
&+{1\over2}\varepsilon_{BN}\bigl(\sqrt2 \chi_{(M}{}^{EFG}H_{X)EFG}\pm 
  D_{EF}H^{EF}{}_{MX}\bigr)+\cr
&+{1\over2}\varepsilon_{BX}\bigl(\sqrt2 \chi_{(N}{}^{EFG}H_{M)EFG}\pm
  D_{EF}H^{EF}{}_{NM}\bigr)+\cr
&+{1\over2}\varepsilon_{BM}\bigl(\sqrt2 \chi_{(X}{}^{EFG}H_{N)EFG}\pm
  D_{EF}H^{EF}{}_{XN}\bigr)\Bigr)
\varepsilon_{RS}\Theta^{MR}_\mu\Theta^{NS}_\nu.
&(3.2.6)\cr}
$$
\ni
Thus the tensors $H_{\mu\nu}$, $B_{\mu\nu}$ have natural twistorial 
interpretation since they represent the nonvanishing components of the 
3-surface twistor curvature. The flatness of the 3-surface twistor 
connection (3.2.3) is therefore equivalent to the vanishing of $H_{\mu
\nu}$ and $B_{\mu\nu}$, i.e., as Tod recognized first, to the local 
isometric imbeddability of $(\Sigma,h_{\mu\nu},\chi_{\mu\nu})$ into some 
conformally flat spacetime. It is an easy calculation to show that under 
spacetime conformal rescalings the spinor expressions above do, in fact, 
transform as the spinor parts of a (1,1) twistor (valued 2-form).\par
    By (3.1.5-6) and (3.2.4-6) the (dual of the) Chern--Simons 3-form 
built from the 3-surface twistor connection ${}^\pm{\tt A}^{\ualpha}_{\mu
{\ubeta}}$ is 

$$\eqalign{
&\varepsilon^{\mu\nu\rho}\Bigl({}^\pm{\tt A}^{\ualpha}_{\mu{\ubeta}}{}^\pm
  {\tt R}^{\ubeta}{}_{{\ualpha}\nu\rho}+{2\over3}{}^\pm{\tt A}^{\ualpha}
  _{\mu{\ubeta}}{}^\pm{\tt A}^{\ubeta}_{\nu{\ugamma}}{}^\pm{\tt A}^{\ugamma}
  _{\rho{\ualpha}}\Bigr)=\cr
&=\varepsilon^{\mu\nu\rho}\Bigl\{\Bigl({}^-\Gamma^{\uA}_{\mu{\uB}}{}^-
  F^{\uB}{}_{{\uA}\nu\rho}+{2\over3}{}^-\Gamma^{\uA}_{\mu{\uB}}{}^-\Gamma
  ^{\uB}_{\nu{\uC}}{}^-\Gamma^{\uC}_{\rho{\uA}}\Bigr)+\Bigl({}^+\Gamma
  ^{\uA}_{\mu{\uB}}{}^+F^{\uB}{}_{{\uA}\nu\rho}+{2\over3}{}^+\Gamma^{\uA}
  _{\mu{\uB}}{}^+\Gamma^{\uB}_{\nu{\uC}}{}^+\Gamma^{\uC}_{\rho{\uA}}\Bigr)
  \Bigr\}=\cr
&=\varepsilon^{\mu\nu\rho}\Bigl\{\Bigl(\Gamma^{\uA}_{\mu{\uB}}F^{\uB}
  {}_{{\uA}\nu\rho}+{2\over3}\Gamma^{\uA}_{\mu{\uB}}\Gamma^{\uB}_{\nu{\uC}}
  \Gamma^{\uC}_{\rho{\uA}}\Bigr)+\overline{\Bigl(\Gamma^{\uA}_{\mu{\uB}}
  F^{\uB}{}_{{\uA}\nu\rho}+{2\over3}\Gamma^{\uA}_{\mu{\uB}}\Gamma^{\uB}_{\nu
  {\uC}}\Gamma^{\uC}_{\rho{\uA}}\Bigr)}\Bigr\}.\cr} \eqno(3.2.7)
$$
\ni
Thus the Chern--Simons functional built from the 3-surface twistor 
connection is just half of the real Sen--Chern--Simons functional, yielding 
a pure twistorial interpretation and a manifest conformally invariant form 
of the latter (and of $Y_{(k,k)}$ for any $k\in{\bf N}$), too. Thus the 
stationary points of the 3-surface twistor Chern--Simons functional, with 
respect to both the 3-surface twistor connection and the fields $h_{\mu
\nu}$ and $\chi_{\mu\nu}$ of the initial data, are just the initial data 
for which the 3-surface twistor connection is flat. Similarly to the higher 
dimensional spinor representations, we expect that the Chern--Simons 
functional built from the 3-surface twistor connections for higher valence 
twistors will essentially coincide with (3.2.7). \par

\bigskip
\bigskip

\ni
{\lbf 4. Time evolution}\par
\bigskip
\ni
{\bf 4.1 Comparison theorem for $Y[\Gamma^{\uA}{}_{\uB}]$}\par
\bigskip
\ni
Next consider a one parameter family of initial data on a fixed 
3-manifold, $(\Sigma,h_{\mu\nu}(t),\chi_{\mu\nu}(t))$, and we ask how the 
Chern--Simons functional, built from the connection ${\cal D}_\mu$ in some 
representation $\rho$, varies as a function of $t$. Obviously, this problem 
can be reinterpreted as the question of its time evolution if $\theta_t:
\Sigma\rightarrow M$, $t\in{\bf R}$, is a foliation of a Lorentzian 
spacetime $(M,g_{ab})$ with spacelike hypersurfaces. On the typical 
hypersurface $\Sigma$ this foliation is represented by a lapse function 
$N(t):\Sigma\rightarrow(0,\infty)$. By (2.3.5) it is enough to 
consider only the time evolution of $Y[\Gamma^{\uA}{}_{\uB}]$. In the 
present subsection we derive a formula by means of which we can compare 
the Chern--Simons functional on two different spacelike hypersurfaces in 
a given globally hyperbolic spacetime. This result is independent of any 
field equation. We take into account Einstein's field equations only in 
the next subsection. \par
      Since $M$ is diffeomorphic to $\Sigma\times{\bf R}$, $M$ admits a 
spinor structure, and let $S^A(M)$ be the (trivializable) spinor bundle 
and $\{\varepsilon^A_{\uA}\}$ a (globally defined) normalized spinor dyad. 
Let $\Sigma_t:=\theta_t(\Sigma)$, $t_a$ its timelike unit normal and let 
$\nabla_e$ be the connection on the spacetime spinor bundle $S^A(M)$. Then 
the connection 1-form of $\nabla_a$ in the dyad $\{\varepsilon^A_{\uA}\}$ 
is ${}^{(4)}\Gamma^{\uA}_{a{\uB}}:=\varepsilon^{\uA}_B\nabla_a\varepsilon^B
_{\uB}$, and the curvature 2-form is ${}^{(4)}R^{\uA}{}_{{\uB}cd}:=
\varepsilon^{\uA}_A\varepsilon^B_{\uB}{}^{(4)}R^A{}_{Bcd}$. Their pull 
back to $\Sigma$ along the imbedding $\theta_t$ are just the connection 
and curvature forms, $\Gamma^{\uA}_{\mu{\uB}}(t)$ and $F^{\uA}{}_{{\uB}\mu
\nu}(t)$, of the Sen connection in the spinor representation at `time' 
$t$, respectively. Thus the pull back to $\Sigma$ of the spacetime 
Chern--Simons 3-form ${}^{(4)}R^{\uA}{}_{{\uB}[ab}{}^{(4)}\Gamma^{\uB}_{c]
{\uA}}+{2\over3}{}^{(4)}\Gamma^{\uA}_{[a\vert{\uB}\vert}{}^{(4)}\Gamma
^{\uB}_{b\vert{\uC}\vert}{}^{(4)}\Gamma^{\uC}_{c]{\uA}}$ along $\theta_t$ 
is the Chern--Simons 3-form built from $\Gamma^{\uA}_{\mu{\uB}}(t)$ on 
$S^A(\Sigma)$. On the other hand, the exterior derivative of the spacetime 
Chern--Simons 3-form, contracted with the volume 4-form, is 

$$
{1\over2}{}^{(4)}R_{{\uA}{\uB}ab}{}^{(4)}R^{{\uA}{\uB}}{}_{cd}\varepsilon
^{abcd}= 4E_{ab}H^{ab}-{\rm i}\Bigl(2\bigl(E_{ab}E^{ab}-H_{ab}H^{ab}\bigr)
-{1\over2}{}^{(4)}G_{ab}{}^{(4)}G^{ab}+{1\over6}{}^{(4)}R^2\Bigr),\eqno(4.1.1)
$$
\ni
where we used the expressions for $E_{ab}$ and $H_{ab}$ given in subsection 
2.2. Thus applying the Stokes theorem to the spacetime Chern--Simons 
3-form on the spacetime domain bounded by the $\Sigma_0$ and $\Sigma_t$ 
hypersurfaces and assuming that the integrals exist, we get 

$$
Y[\Gamma^{\uA}{}_{\uB}(t)]-Y[\Gamma^{\uA}{}_{\uB}(0)]=\int^t_0\int_{\Sigma
_{t'}}{1\over2}{}^{(4)}R_{{\uA}{\uB}ab}{}^{(4)}R^{{\uA}{\uB}}{}_{cd}
\varepsilon^{abcd}\,N\,{\rm d}\Sigma_{t'}\,{\rm d}t'.\eqno(4.1.2)
$$
\ni
Note that by (4.1.2) we can compare $Y[\Gamma^{\uA}{}_{\uB}]$ on any two, 
maybe intersecting, hypersurfaces $\Sigma'$ and $\Sigma''$ if there is a 
hypersurface $\Sigma$ and there is a foliation $\Sigma'_t$ between 
$\Sigma'$ and $\Sigma$ and a foliation $\Sigma''_t$ between $\Sigma''$ and 
$\Sigma$. The real part of the right hand side in (4.1.1) transforms as 
$4E_{ab}H^{ab}\mapsto 4\Omega^{-4}E_{ab}H^{ab}$ under spacetime conformal 
rescalings. Thus, recalling that by its very definition, $Nt^a\nabla_at=1$, 
the lapse function $N$ transforms as $N\mapsto\Omega N$, {\it the time 
derivative of the Chern--Simons functionals defined in the tensor 
representations is also conformally invariant}. In general neither the real 
nor the imaginary part has definite sign. If however the scalar polynomial 
invariant $I:=\psi_{ABCD}\psi^{ABCD}={1\over8}C_{abcd}(C^{abcd}+{\rm i}\ast 
C^{abcd})=E_{ab}E^{ab}-H_{ab}H^{ab}+{\rm i}2E_{ab}H^{ab}$ of the spacetime 
Weyl tensor is real, then the {\it real} Sen--Chern--Simons functional is 
not only a conformal invariant of the initial data set but, being 
independent of the initial hypersurface, a {\it conformal invariant of the 
whole spacetime}. This holds for a large class of (algebraically general 
and special) spacetimes, including all Petrov III. and N spacetimes [20]. 
If the invariant $I$ itself is zero and in addition the Ricci tensor is also 
vanishing then the Ashtekar--Chern--Simons functional is also independent 
of the initial hypersurface and provides a further {\it global invariant 
of the whole spacetime}. \par
\bigskip

\bigskip
\ni
{\bf 4.2 On the time evolution via Einstein's equations, on a `natural 
time variable' and examples}\par
\bigskip
\ni
Instead of the Stokes theorem of the previous subsection we could start 
with a choice for the lapse and shift and could evolve $Y[\Gamma^{\uA}{}
_{\uB}(t)]$ in time by using (2.3.3) and evolving the initial data via the 
3+1 form of Einstein's equations. What we would obtain, however, is just 
(the time derivative of) (4.1.2) if in which the Einstein equations 
${}^{(4)}G_{ab}+\Lambda g_{ab}=-\kappa T_{ab}$ had already been taken into 
account: The terms containing the shift integrate to zero because of the 
diffeomorphism invariance of $Y_\pm$, furthermore, although in general the 
conformal electric curvature $E_{\mu\nu}$ cannot be expressed by the 
intrinsic and extrinsic geometrical data on $\Sigma$, by Einstein's 
equations it {\it becomes} an expression of the geometric data {\it and} 
the energy-momentum tensor on $\Sigma$. Explicitly $E_{ab}=-(R_{ab}+V_{ab}
-{1\over3}h_{ab}(R+V))-{1\over2}\kappa(\sigma_{ab}-{1\over3}\sigma^e{}_e 
h_{ab})$, where we used the standard 3+1 decomposition, $T_{ab}=\mu t_at_b+
2J_{(a}t_{b)}+\sigma_{ab}$, of the energy-momentum tensor. \par
     Recently the imaginary part of the Ashtekar--Chern--Simons functional 
as a natural internal time variable was suggested in the {\it configuration 
space} of cosmological models [21]. Then the question arises as whether one 
can introduce a natural {\it internal} time variable in the {\it spacetime}, 
at least with respect to a foliation $\theta_t$, given geometrically in 
cosmological spacetimes. This would have to be monotonic with respect to 
the coordinate time $t$ above. (For such earlier suggestions see, for 
example, [22,23].) In fact, for the $t={\rm constant}$ hypersurfaces of 
the maximal spacelike symmetry of the closed Robertson--Walker metrics 
(4.1.1) is 
$-6{\rm i}a^{-3}\ddot a(\dot a^2+1)$, where by Einstein's equations the 
scale function $a(t)$ satisfies $3a^{-1}\ddot a=\Lambda-{1\over2}\kappa
(\mu+3p)$ and $3(\dot a^2+1)=a^2(\kappa\mu+\Lambda)$, and the pressure $p$ 
is defined by $\sigma_{ab}=:-ph_{ab}$. Therefore ${\rm Im}Y_\pm$ {\it is} 
in fact monotonic if the strong energy condition is satisfied and e.g. 
$\Lambda\leq0$. (For a discussion of the role of the cosmological constant 
in the dynamics of the Robertson--Walker spacetimes in general see e.g. 
[24].) It would therefore be 
interesting to see $Y[\Gamma^{\uA}{}_{\uB}]$ itself. We calculate this by 
calculating the spinor Chern--Simons invariant for the more general 
homogeneous Bianchi cosmological spacetimes using the technique of [25,26], 
where the 3-space is still assumed to be a group manifold. To ensure 
the existence of the integral of the Chern--Simons 3-form we must assume 
that the typical Cauchy hypersurfaces are compact. The only three compact 3 
dimensional manifolds admitting Lie group structure are the torus $S^1
\times S^1\times S^1$, the 3-sphere $S^3\approx SU(2)$ and the projective 
space $S^3/Z_2\approx SO(3)$, for which the structure constants can be 
written as $c^{\bk}_{\bi\bj}=c\eta^{\bk\bl}2\varepsilon_{\bl\bi\bj}$. Here 
$c=0$ for the torus and $c=1$ for $SU(2)$ and $SO(3)$. Since from the point 
of view of the Chern--Simons functional the only difference between the 
$SU(2)$ and $SO(3)$ cases is that the integration domain for $SO(3)$ is 
half of that for $SU(2)$, we calculate $Y[\Gamma^{\uA}{}_{\uB}]$ only for 
the torus and the sphere. Thus let $\{\sigma^{\bi}_\mu\}$ be a left 
invariant 1-form basis in which the structure constants are the ones given 
above, i.e. ${\rm d}\sigma^{\bi}=\eta^{\bi\bj}c\epsilon_{\bj\bk\bl}\sigma
^{\bk}\wedge\sigma^{\bl}$. (With the choice $\eta^{\bk\bl}2\varepsilon_{\bl
\bi\bj}$ for the structure constant on $SU(2)$ the basis $\{\sigma^{\bi}
_\mu\}$ will be ortho{\it normal} with respect to the unit sphere metric 
inherited from ${\bf R}^4$ through the canonical imbedding.) Let $h_{\bi
\bj}$ and $\chi_{\bi\bj}$ be the components of the metric and extrinsic 
curvature in the basis $\{\sigma^{\bi}_\mu\}$, respectively, and $h^{\bi
\bj}$ the inverse and $h$ the determinant of $h_{\bi\bj}$. Then a direct 
calculation yields 

$$\eqalign{
Y[\Gamma^{\uA}{}_{\uB}]=-{{\rm i}\over3}&\Bigl\{\bigl(\chi_{\bi\bj}h^{\bi
  \bj}\bigr)^3-3\bigl(\chi_{\bi\bj}h^{\bi\bj}\bigr)\bigl(\chi_{\bk\bl}h
  ^{\bl\bm}\chi_{\bm\bn}h^{\bn\bk}\bigr)+2\chi_{\bi\bj}h^{\bj\bk}\chi_{\bk
  \bl}h^{\bl\bm}\chi_{\bm\bn}h^{\bn\bi}\Bigr\}\sqrt{\vert h\vert}{\rm Vol}_0
  (\Sigma)-\cr
-{\rm i}c{4\pi^2\over\sqrt{\vert h\vert}}\chi_{\bi\bj}&\Bigl\{-8\eta^{\bj
  \bk}h_{\bk\bl}\eta^{\bl\bj}+4\eta^{\bi\bj}\bigl(h_{\bk\bl}\eta^{\bk\bl}
  \bigr)+2h^{\bi\bj}\bigl(h_{\bk\bl}\eta^{\bk\bm}\eta^{\bl\bn}h_{\bm\bn}
  \bigr)-h^{\bi\bj}\bigl(h_{\bk\bl}\eta^{\bk\bl}\bigr)^2\Bigr\}-\cr
-2\pi^2c&\Bigl\{6\chi_{\bi\bj}h^{\bj\bk}\chi_{\bk\bl}\eta^{\bl\bi}+\bigl(
  h_{\bi\bj}\eta^{\bi\bj}\bigr)\Bigl(\bigl(\chi_{\bk\bl}h^{\bk\bl}\bigr)^2-
  \chi_{\bk\bl}h^{\bk\bm}h^{\bl\bn}\chi_{\bm\bn}\Bigr)-4\bigl(\chi_{\bi\bj}
  h^{\bi\bj}\bigr)\bigl(\chi_{\bk\bl}\eta^{\bk\bl}\bigr)-\cr
&-{4\over3\vert h\vert}\Bigl[10h_{\bi\bj}\eta^{\bj\bk}h_{\bk\bl}\eta^{\bl
  \bm}h_{\bm\bn}\eta^{\bn\bi}-9\bigl(h_{\bi\bj}\eta^{\bi\bj}\bigr)\bigl(h
  _{\bk\bl}\eta^{\bl\bm}h_{\bm\bn}\eta^{\bn\bk}\bigr)+2\bigl(h_{\bi\bj}\eta
  ^{\bi\bj}\bigr)^3\Bigr]\Bigr\}.\cr}\eqno(4.2.2)
$$
\ni
Here ${\rm Vol}_0(\Sigma):=\int_\Sigma\sigma^1\wedge\sigma^2\wedge\sigma^3$, 
the (non-dynamical) left invariant group volume, which is $2\pi^2$ for $c=
1$ and it is chosen to be $8\pi^3$ for $c=0$. \par
       The general $c=0$ {\it vacuum} solution is $ds^2=(dt)^2-t^{2p_1}
(dx^1)^2-t^{2p_2}(dx^2)^2-t^{2p_3}(dx^3)^2$, where $\sigma^{\bi}_\mu=D_\mu 
x^{\bi}$ and $x^{\bi}\in[0,2\pi]$, and $p_1+p_2+p_3=1$, $p^2_1+p^2_2+p^2_3
=1$ (the spatially closed Kasner solution [20]). Then (4.2.2) gives 
$Y[\Gamma^{\uA}{}_{\uB}]=-{16\over3}\pi^3{\rm i}(p_1^3+p_2^3+p_3^3-1)$, a 
purely {\it imaginary constant}. Thus this is an algebraically general 
spacetime for which the spinor Chern--Simons functional is an invariant 
for the whole spacetime, and, in particular, ${\rm Im}\,Y_\pm$ is certainly 
{\it not} a time function. This result shows that the generalization of 
the (Riemannian) conformal invariant $Y[h_{\mu\nu}]$ of Chern and Simons 
for initial data sets is not trivial: $Y_\pm[h_{\mu\nu},\chi_{\mu\nu}]$ 
depends essentially on $\chi_{\mu\nu}$, and, apart from the permutations 
of the Kasner exponents $(p_1,p_2,p_3)$, characterizes the vacuum Bianchi 
I. cosmological spacetimes completely. (To see this it may help the use of 
the {\it global} explicit parameterization $3p_1=1-\cos\alpha+\sqrt{3}
\sin\alpha$, $3p_2=1-\cos\alpha-\sqrt{3}\sin\alpha$, $3p_3=1+2\cos\alpha$, 
where $\alpha\in[0,2\pi]$.) 
For $c=1$ the left invariant basis may be chosen such that $2\sigma^1_\mu= 
\sin\psi D_\mu\theta-\cos\psi\sin\theta D_\mu\phi$, $2\sigma^2_\mu=\cos
\psi D_\mu\theta+\sin\psi\sin\theta D_\mu\phi$ and $2\sigma^3_\mu=-D_\mu\psi
-\cos\theta D_\mu\psi$, where $\psi\in[0,4\pi]$, $\phi\in[0,2\pi]$, $\theta 
\in[0,\pi]$ are the standard Euler angle coordinates on the 3-sphere. Then 
for the Robertson--Walker line element we get $Y[\Gamma^{\uA}{}_{\uB}]=8
\pi^2-4\pi^2{\rm i}\dot a(\dot a^2+3)$. Thus the conformal invariant $Y_0$ 
for any closed Robertson--Walker initial data set is zero, but ${\rm 
Im}\,Y_\pm$ is monotonic provided, as we saw, the strong energy condition 
is satisfied. 
A non-isotropic solution with the stiff equation of state, $p=\mu$, and 
vanishing cosmological constant was found by Barrow [27]. It has the form 
$ds^2=(dt)^2-a^2(t)(\sigma^3)^2-b^2(t)((\sigma^1)^2+(\sigma^2)^2)$ with 
the scale functions given by $a^2(\tau)=4A\,{\rm sech}(A\tau)$ and $b^2
(\tau)=A^{-1}B^2\cosh(A\tau){\rm sech}^2({1\over2}B(\tau+\tau_0))$, where 
$A$, $B$ and $\tau_0$ are constants and the parameter $\tau$ is defined 
implicitly by ${{\rm d}t\over{\rm d}\tau}={1\over8}a(\tau)b^2(\tau)$. Then 
the energy density is $\kappa\mu=16(B^2-A^2)a^{-2}b^{-4}$ and the metric 
becomes isotropic if $a=b$, whenever $B=2A$ (and $\tau_0=0$). For the $t=
{\rm const.}$ hypersurface in this spacetime we find 

$$\eqalign{
Y[\Gamma^{\uA}{}_{\uB}]&=8\pi^2\Bigl\{1+\Bigl({a^2-b^2\over b^2}\Bigr)^2+
  {1\over4b^4}\Bigl(\sqrt{16A^2-a^4}-\sqrt{4B^2-a^2b^2}\Bigr)^2\Bigr\}+\cr
&+{4\pi^2{\rm i}\over a^2b^4}\sqrt{16A^2-a^4}\Bigl(\sqrt{16A^2-a^4}-
  2\sqrt{4B^2-a^2b^2}\Bigr)^2+\cr
&+{16\pi^2{\rm i}\over b^2}\Bigl(\bigl(1-{a^2\over4b^2}-{b^2\over a^2}
  \bigr)\sqrt{16A^2-a^4}+{a^2\over b^2}\sqrt{4B^2-a^2b^2}\Bigr).\cr}
$$
\ni
Therefore the anisotropy of the geometry, characterized by the differences 
$\sqrt{16A^2-a^4}-\sqrt{4B^2-a^2b^2}$ and $a^2-b^2$ of the extrinsic 
curvature and the intrinsic metric, respectively, contribute both to the 
real and imaginary parts of $Y_\pm$, too, i.e. $Y_0$ is also a non-trivial 
generalization of the Riemannian $Y[h_{\mu\nu}]$. Both the real and 
imaginary parts of $Y[\Gamma^{\uA}{}_{\uB}]$ are changing in time, but, 
in general, ${\rm Im}\,Y[\Gamma^{\uA}{}_{\uB}]$ is {\it not} monotonic 
even if the energy condition $A^2<B^2$ is assumed to hold. \par
      Returning to the general formula (4.1.1), neither its real nor its 
imaginary part has definite sign even if Einstein's equations are taken 
into account and the dominant energy condition is satisfied. Moreover 
the indefinite expression $E_{ab}E^{ab}-H_{ab}H^{ab}$ is still present 
even in vacuum. Since however for the analogous expression in Ricci-flat 
Riemannian 4-geometries we get $E_{ab}E^{ab}+H_{ab}H^{ab}$, which is 
positive definite (and in a Lorentzian spacetime this would be just the 
well known Bel--Robinson tensor contracted with the unit normal $t^a$), 
one could hope to obtain positive definite imaginary term for the Wick 
rotated Ashtekar, i.e. the Barbero connection [7,8]. In fact, the same 
analysis can be repeated for the Barbero connection, obtaining a formula 
analogous to (4.1.1). But that is much more complicated and neither the 
real nor the imaginary part appears to have a definite sign either. Hence 
${\rm Im}\,Y_\pm$ can be interpreted as a natural time variable in the 
spacetime only for a very limited class of cosmological models. \par
      Finally, to have an asymptotically flat example for the Chern--Simons 
functional, we can compute $Y[\Gamma^{\uA}{}_{\uB}]$ for the 
Reissner--Nordstr\"om spacetime. For the 3-manifold $\Sigma$ we choose 
a maximally extended $t={\rm const.}$ spacelike hypersurface, consisting 
of two asymptotically flat `ends' and joining together at the surface 
of bifurcation of the event horizon. It is easy to find a globally defined 
triad $\{e^\mu_{\bi}\}$ on $\Sigma$, using the fact that $\Sigma$ with the 
induced metric $h_{\mu\nu}$ is globally conformally flat (see e.g. [28]). 
However, just because of its global conformal flatness and the fact that 
this hypersurface is time symmetric, the whole $Y[\Gamma^{\uA}{}_{\uB}]$ 
is zero. The analogous calculation for the Kerr--Newman solution would be 
much more complicated. \par

\bigskip
\bigskip

\ni
{\lbf Acknowledgements}\par
\bigskip
\ni
I am grateful to Robert Beig for the numerous useful discussions and, in 
particular, for the idea of using the Stokes theorem instead of Einstein's 
equations in Section 4. Thanks are due to J\"org Frauendiener and George 
Sparling for comments and notes, and to John Wainwright for pointing out 
the Barrow solution (Ref. [27]). This work was partially supported by the 
Hungarian Scientific Research Fund grants OTKA T016246 and OTKA T030374. 

\bigskip
\bigskip
\ni
{\lbf References:}\par
\bigskip

\item{[1]}  S.S. Chern, J. Simons, {\it Characteristic forms and geometrical 
            invariants}, Ann. Math. {\bf 99} 48 (1974)
\item{[2]}  R. Beig, L.B. Szabados, {\it On a global conformal invariant 
            of initial data sets}, Class. Quantum Grav. {\bf 14} 3091 (1997) 
\item{[3]}  R. Meyerhoff, {\it Hyperbolic 3-manifolds with equal volumes but 
            different Chern--Simons invariants}, in Low Dimensional Topology 
	     and Kleinian Groups, Ed.: D.B.A. Epstein, Lecture Notes No 112, 
	     Cambridge University Press 1986
\item{[4]}  H. Friedrich, {\it Einstein equations and conformal structure: 
            Existence of anti-de-Sitter-type spacetimes}, J. Geom. Phys. 
	     {\bf 17} 125 (1995)
\item{[5]}  H. Friedrich, {\it Gravitational fields near space-like and null 
            infinity}, J. Geom. Phys. {\bf 24} 83 (1998) 
\item{[6]}  K.P. Tod, {\it Three surface-twistors and conformal embedding}, 
            Gen. Rel. Grav. {\bf 16} 435 (1984)
\item{[7]}  J.F. Barbero, {\it Real Ashtekar variables for Lorentzian 
            signature spacetimes}, Phys.Rev.D {\bf 51} 5507 (1995)
\item{[8]}  G. Immirzi, {\it Real and complex connections for canonical 
            gravity}, Class. Quantum Grav. {\bf 14} L177 (1997)
\item{[9]}  R. Penrose, W. Rindler, {\it Spinors and spacetime}, vol 1, 
            Cambridge University Press, Cambridge, 1982 
\item{   }  R. Penrose, W. Rindler, {\it Spinors and spacetime}, vol 2, 
            Cambridge University Press, Cambridge, 1986
\item{   }  S.A. Hugget, K.P. Tod, {\it An introduction to twistor theory}, 
            Cambridge University Press, Cambridge 1985
\item{[10]} S. Kobayashi, K. Nomizu, {\it Foundation of differential 
            geometry}, vol 1, Interscience, New York 1964 
\item{   }  S. Kobayashi, K. Nomizu, {\it Foundation of differential 
            geometry}, vol 2, Interscience, New York 1968
\item{[11]} M. Spivak, {\it A comprehensive introduction to differential 
            geometry}, vol 1, Publish or Perish, Houston, TX 1979
\item{[12]} A. Sen, {\it On the existence of neutrino `zero modes' in 
            vacuum spacetimes}, J. Math. Phys. {\bf 22} 1781 (1981)
\item{[13]}  P. Sommers, {\it Space spinors}, J. Math. Phys. {\bf 21} 2567 
            (1980)
\item{[14]} O. Reula, {\it Existence theorem for solutions of Witten's 
            equation and nonnegativity of total mass}, J. Math. Phys. {\bf 
	     23} 810 (1982)
\item{[15]} J. Frauendiener, {\it Triads and the Witten equation}, Class. 
            Quantum Grav. {\bf 8} 1881 (1991)
\item{[16]} R. Geroch, S.-M. Perng, {\it Total mass-momentum of arbitrary 
            initial data sets in general relativity}, J. Math. Phys. {\bf 35} 
	     4157 (1994)
\item{[17]} L.B. Szabados, {\it Two dimensional Sen connections in general 
            relativity}, Class. Quantum Grav. {\bf 11} 1833 (1994)
\item{[18]} J. Lewandowski, {\it Twistor equation in a curved spacetime}, 
            Class. Quantum Grav. {\bf 8} L11 (1991)
\item{[19]} K.P. Tod, {\it Some examples of Penrose's quasi-local mass 
            construction}, Proc.Roy.Soc.Lond. A {\bf 388} 457 (1983)
\item{[20]} D. Kramer, H. Stephani, M.A.H. MacCallum, E. Herlt, {\it Exact 
            solutions of Einstein's Field equations}, Cambridge University 
            Press, Cambridge 1980
\item{[21]} L. Smolin, C. Soo, {\it The Chern--Simons invariant as the 
            natural time variable for classical and quantum cosmology}, 
            Nucl. Phys. B. {\bf 449} 289 (1995)
\item{[22]} C.W. Misner, Mixmaster universe, Phys.Rev.Lett. {\bf 22} 1071 
            (1969)
\item{[23]} J.W. York, Role of conformal three-geometry in the dynamics of 
            gravitation, Phys.Rev.Lett. {\bf 28} 1082 (1972)
\item{[24]} S.W. Hawking, G.F.R. Ellis, {\it The large scale structure of 
            spacetime}, Cambridge University Press, Cambridge 1973
\item{[25]} M.A.H. MacCallum, Anisotropic and inhomogeneous relativistic 
            cosmologies, in {\it General relativity, An Einstein
            centenary survey}, Ed.: S.W. Hawking, W. Israel, Cambridge
            University Press, Cambridge 1979,
\item{[26]} R.M. Wald, {\it General relativity}, University of Chicago 
            Press, Chicago, 1986
\item{[27]} J.D. Barrow, Quiescent cosmology, Nature {\bf 272} 211 (1978)
\item{[28]} G. Gibbons, The isoperimetric and Bogomolny inequalities for 
            black holes, in {\it Global Riemannian Geometry}, Eds.: T.J. 
	    Willmore, N. Hitchin, Ellis Horwood Ltd, New York 1984

\end